\documentstyle[12pt]{article}
\catcode`@=11
\@addtoreset{equation}{section}
\catcode`@=12
\begin{document}
\title{Hamiltonian BFV--BRST
theory of closed quantum cosmological models} \author{A. Yu.
Kamenshchik\thanks{Permanent address:
Nuclear Safety Institute, Russian
Academy of Sciences, 52 Bolshaya Tulskaya, Moscow, 113191, Russia.
Electronic address: grg@ibrae.msk.su}
 \, and \ S. L. Lyakhovich\thanks{Permanent address:
Theoretical Physics Department, Tomsk State University, pr.
Lenin 36, Tomsk 634050, Russia. Electronic address: sll@phys.tsu.tomsk.su}}
\date{}
\maketitle
\begin{center}
{\it International Centre for Theoretical Physics,}
\end{center}
\begin{center}
{\it P.O. BOX 586, Trieste 34100 , Italy}
\end{center}

\begin{abstract}
We introduce and study a new discrete basis of gravity constraints
by making use of harmonic expansion for closed cosmological models.
The full set of constraints is splitted into area-preserving
spatial diffeomorphisms, forming closed subalgebra, and Virasoro-like
generators. Operatorial Hamiltonian BFV-BRST quantization is performed
in the framework of perturbative expansion in the dimensionless
parameter which is a positive power of the ratio of Planckian volume
to the volume of the Universe. For the $(N+1)$ - dimensional
generalization of stationary closed Bianchi-$I$ cosmology the
nilpotency condition for the BRST operator is examined in the first
quantum approximation. It turns out, that certain relationship
between dimensionality of the space and the spectrum of matter
fields emerges from the requirement of quantum consistency of the model.
\end{abstract}



\section{Introduction}

At the present time we observe the explosive development in quantum 
gravity and string theory which have many-sided interconnections 
\cite{strings}. 
From the standpoint of the constrained dynamics 
\cite{constrained} these theories manifest a certain similarity 
originating from the reparametrization invariance which is a gauge 
symmetry of the both.
Meanwhile, the theories are usually
treated in an essentially different way 
in respect to the perturbative expansions. Various well-advanced 
string models, having a trend to be thought about as dual ones, are 
considered nowadays as 
expansions around different limiting values of a coupling constant of 
an unknown fundamental superstring theory  \cite{strings1}.
Oppositely, gravity has a widely recognized candidate for a 
fundamental theory that is Einstein action or, perhaps, its some or
another extension, but the perturbative
treatment is a longstanding unsolved problem of the General Relativity.

The main intention of the present work is to undertake an attempt of 
the Hamiltonian BFV--BRST quantization of closed cosmological models by
exploiting the keenship between the string and gravity gauge 
symmetries.  

Apparently,
a phase-space constrained description is very similar for string
theory and gravity while they are being carried out locally with dynamical
variables and constraints defined at a point of a space-like section of
world-sheet or spacetime, respectively. However, there is an essential
distinction in habitual approaches to quantization of strings and gravity.

Quantizing strings, one may work with a discrete set of modes,
describing string excitations, and a discrete set of constraints, which
form
Virasoro algebra. In doing so, one observes that quantum commutators of
the constraints acquire central extension, and the quantum consistency
of the theory is provided for the critical values of dimensionality and
intercept. One of the most efficient methods to observe this phenomenon
is the BFV-BRST Hamiltonian quantization of constrained systems
\cite{BFV-BRST,BF-operator,BF-86,Bat-spur}.
In the framework of this method one can show how the
nilpotency condition for the BRST charge defines the values of
critical parameters for strings \cite{Hwang,Okhta}, membranes
\cite{membranes} and $W_{3}$ gravity \cite{Tierry}.
Actually, the use of discrete
set of modes and constraints is merely a convenient tool for taking
into account the topology of the compact world-sheet spacelike section,
when the method is applied.

Canonical quantization of gravity and cosmology
is usually treated by either perturbative consideration of path
integral or operatorial quantization according to Dirac scheme with
constraints imposed onto physical states. Basically, such
a treatment implements continuous basis of dynamical modes and
constraints, which can satisfactory describe the dynamics locally
but are hardly
well-defined globally on compact spacelike sections of the spacetime.
Meanwhile, the consideration of cosmological perturbations
in terms of globally defined harmonics was used for the investigation
of different problems in classical \cite{Khalat} and quantum
\cite{Hal-Haw,Bar-Kam} cosmology. However, a globally defined discrete
constraint basis for quantum cosmology of closed Universes has not been
hitherto discussed.
Consistent canonical quantization requires both the dynamical modes
and constraints to be globally defined,
since this is essential for the adjustment of the ordering of the
modes to the quantum algebra of the constraints. For example,
in the case of the Wick ordering of bosonic string modes, the
quantum corrections to the constraint commutators give rise to the central
extension of the Virasoro algebra representation in the Fock space. Then
the quantum constraint algebra can be made consistent provided the proper
basis of constraints is chosen ($L$ and $\bar{L}$ are employed instead
of habitual for cosmologists lapse and shift generators $H_{\perp}, \ H_{||}$
and moreover, only one half of $L_n$ and $\bar{L}_m$ should
annihilate physical states). At the mean time,
the straightforward pursuing
the Dirac scheme, with $H_{\perp}$ and $H_{||}$ being imposed on the
physical subspace, leads to a contradiction.

In this paper, we suggest a new scheme for canonical quantization of
the closed cosmological models. The distinguishing features of this
treatment are as follows: the decomposition of both dynamical
variables and constraints into harmonics which are the eigenfunctions
of the Laplace operator for the maximally symmetric space of given
topology; separation of subalgebra of area-preserving diffeomorphisms
from a total set of gravity constraints whereas the rest of the
constraints form Virasoro-like generators; quantization of this
constrained theory is performed in the framework of the Hamiltonian
BFV-BRST formalism with due regard to the ordering of gravity,
matter and ghost harmonics; studying the quantum nilpotency condition
for the BRST operator, we apply a certain perturbative expansion of the
constraints and structure functions with a small parameter ${l_P}/V^{1/N}$,
where $l_P$ is a Planck length and $V$ is a spatial volume of the Universe,
$N$ is a dimensionality of the space. Implementing this scheme,
we focus the consideration
mostly at the stationary Bianchi-$I$ type Universe, although any other closed
cosmological model could be treated in a similar way. The specificity of a
particular topology is encoded in the constraint algebra structure functions
which appear to be expressed via Clebsch--Gordan coefficients of the
corresponding symmetry group representation. In the case of Bianchi-$I$
cosmology, these coefficients are the simplest,
as the group is ${U(1)}^N$. Although we
take this particular
type of a closed cosmological model for the sake of technical
simplicity, it is, however, of a certain physical interest \cite{Bianchi-$I$}.
On the other hand, $N$--torus allows to exploit a straightforward analogy
with the closed bosonic string sigma--model when the nilpotency of the
quantum BRST charge is examined. The most striking outcome of this quantization
procedure, being applied to the Einstein gravity coupled to the matter fields,
is that the number of the matter degrees of freedom appears to be
correlated with the
dimensionality of the space $N$. For example, when there are $d$ mass less
scalar fields only, already the first quantum correction
gives rise to the relation between $d$ and $N$:
$$
d = 30 + \frac{5}{2}(N+1)(N-2),
\eqno(1.1)
$$
which is a necessary condition for nilpotency of the quantum BRST charge.
Mention the curious fact: if one puts $N=1$ (thereby torus reduces
to a circle and the constraints form the Virasoro algebra)
then relation (1.1)
results in $d=25$. This result can be naturally understood from the standpoint
of the string sigma--model without Weyl invariance
\cite{BFLP}. As a matter of fact, bosonic string theory should not be thought
about as one-dimensional limit of the Einstein gravity coupled to a set
of $d$ scalar fields, because string possesses an extra gauge symmetry,
namely Weyl invariance, besides the diffeomorphisms, whereas the mentioned
$\sigma$-model does not have Weyl invariance. As is known \cite{BFLP} the
critical dimensionality of the $\sigma$-model is 25, in contrast with 26
 for strings. The discrepancy in these dimensionalities can be easily
 explained if one  remembers that Weyl invariance gauges out conformal
 mode of 1+1-metric, while in the non-conformal $\sigma$-model case this
 mode contributes to the Virasoro generators on an equal footing with string
 excitations \cite{BFLP}.
 Thus, in some sense, the critical dimensionality for the both
 cases is 26, but for the $\sigma$-model this number consists of
 25 string coordinates and 1 gravity conformal mode.

In its basic features the suggested scheme should, probably, be actual
for other closed cosmological models, including those with non-
stationary classical background. But the critical relations analogous
to (1.1) may change their form for each particular model.

The structure of the paper is as follows: in the second section we describe
a general decomposition of the gravity constraints into the discrete set
of area-preserving diffeomorphisms and Virasoro-like generators and sketch
the way of the representing the structure constants of the constraint
algebra via Clebsch-Gordan coefficients of the corresponding symmetry group;
in the third section we suggest the split of integer vectors on $N$-torus into
two classes which furnishes a means of choosing the Wick ordering for a
proper
part of dynamical modes and constraints; in sec. 4 we write down explicitly
the decomposition of gravity constraints in new discrete basis and the
corresponding structure constants for closed $N$-dimensional Bianchi-$I$
cosmological model ($N$-torus); in sec. 5 we define perturbative expansion
of the constraint operators and calculate first quantum corrections to
the constrained algebra; sec. 6 is devoted to the investigation
of the structure of BRST operator and examination of its nilpotency
condition, as a result, we come to the critical relation between the
dimensionality of the space and the spectrum of matter fields;
sec. 7 contains conclusions.

\section{Gravity constraint algebra for a closed Universe. \newline
Harmonic expansion, \newline
area-preserving diffeomorphisms \newline
and Virasoro-like generators}

Hamiltonian description of any reparametrization-invariant theory
with metric, being considered
as a dynamical variable, contains a set of first-class
phase-space constraints $H_{\perp}$ and $H_{i}, \, i=1,\ldots,N$ where
$N$ is a dimensionality of space. $H_{\perp}$, usually called
super-Hamiltonian or Wheeler-DeWitt operator
(the canonical Hamiltonian vanishes on the constraint surface
for this case), generates lapse
transformations of spacelike hypersurface, while $H_{i}$, called supermomentum,
generates spatial diffeomorphisms \cite{geometrodynamics}. If the gravity
action does not contain higher derivatives of metric, these constraints
can be represented in the following form
\begin{equation}
H_{\perp} = l_{P}^{N-1} G_{ij,kl}\pi^{ij}\pi^{kl} - \frac{1}
{l_{P}^{N-1}} g R + H_{\perp\;matter},
\label{superhamiltonian}
\end{equation}
\begin{equation}
H_{i} = -2g_{ij} \pi^{jk},_{k} - (g_{ik,m} + g_{im,k} - 
g_{km,i})\pi^{mk} + H_{i\;matter},
\label{supermomentum}
\end{equation}
where $g_{ij}$ is a metric on a spacelike section of spacetime,
$\pi^{ij}$ is a momentum conjugated to metric and representing
 extrinsic
curvature of the spacelike hypersurface, $l_{P}$ is the Planck length.
The DeWitt supermetric $G_{ij,kl}$ has the following form:
\begin{equation}
G_{ij,kl} = \frac{1}{2} \left(g_{ik} g_{jl} + g_{il} g_{jk} 
-\frac{2}{N-1} g_{ij} g_{kl}\right),
\label{supermetric}
\end{equation}
while its inverse looks as
\begin{equation}
G^{ij,kl} = \frac{1}{2}\left(g^{ik} g^{jl} + g^{il} g^{jk} - 
2g^{ij}g^{kl}\right).
\label{supermetric1}
\end{equation}

$H_{\perp\;matter}$ and $H_{i\;matter}$ are the matter field contributions
to super-Hamiltonian and supermomenta, respectively. Their explicit
form depends upon a specific set of the matter fields and their
interactions, although the involution relations between the constraints
are not effected by the structure of matter part of these expressions,
because these relations have a purely geometric origin \cite{Teit}.
These involution relations are
\begin{eqnarray}
&&\{H_{\perp}(x),H_{\perp}(x')\}
= g(x)g^{ij}(x)H_{i}(x)\frac{\partial}{\partial x^{j}}
\delta(x,x')\nonumber \\
&&- g(x')g^{ij}(x')H_{i}(x')\frac{\partial}{\partial x^{'j}}
\delta(x,x'),
\label{involution1}
\end{eqnarray}
\begin{eqnarray}
&&\{H_{i}(x),H_{\perp}(x')\} = H_{\perp}(x)\frac{\partial}
{\partial x^{i}} \delta(x,x')\nonumber \\
&&-H_{\perp}(x')\frac{\partial}
{\partial x^{'i}} \delta(x,x'),
\label{involution2}
\end{eqnarray}
\begin{eqnarray}
&&\{H_{i}(x),H_{j}(x')\} = H_{j}(x)\frac{\partial}{\partial x^{j}}
\delta(x,x')\nonumber \\
&&-H_{i}(x')\frac{\partial}{\partial x^{'j}}
\delta(x,x').
\label{involution3}
\end{eqnarray}
Here $\{,\}$ denotes conventional Poisson brackets. The expression for
super-Hamiltonian (\ref{superhamiltonian}) differs from the conventional
one by the factor $g^{1/2}$, thereby the involution relations
(\ref{involution1}),(\ref{involution2}) also have distinctions from
the commonly used. Such an unusual definition of super-Hamiltonian
(which, however, sometimes occurs \cite{FaddeevPopov}) is convenient
for our purposes.
Notice, that the involution relations (\ref{involution1})-
(\ref{involution3}) could not be reproduced
without gravity dynamical variables, because in its essence they express
the gravity gauge symmetry algebra in phase space
\cite{Hojman}.

Considering compact manifolds, one can express phase space variables and
constraints among them via a discrete set of coefficients of harmonic
expansion. The functional basis of the expansion is formed by eigenfunctions
of the Laplace operators defined in the maximally
symmetric space of a topology
under consideration. For the case of Bianchi$I$ cosmology these functions are
simply elements of Fourier expansion on a torus, for the case of Bianchi$IX$
cosmology -- hyperspherical functions \cite{Khalat}. From the viewpoint
of local consideration this harmonic representation is completely equivalent
to the original continuous parameterization of the phase space and constraints,
but the benefit of this treatment is that the expansion, being well-defined on
the manifold, reflects explicitly all the topological distinctions between
different types of cosmological models. The significance of this topological
specificity becomes obvious at comparing open and closed bosonic strings.
Locally, both the models are described by the same canonical variables and
constraints, whereas the difference transparently manifests itself in the
spectrum of modes and constraints upon the Fourier expansion. Really, the
dynamical modes of closed string are left-- and right--moving excitations
$a_n^{\nu}, \bar{{a_m^{\mu}}}$ subjected to two copies of Virasoro constraints
$L_n$ and $\bar{L_m}$, while in the case of the open string the dynamical modes
are the standing waves with a single set of Virasoro constraints $L_n$.

Now, we shall try to push the consideration of the constrained Hamiltonian
formalism along the lines of harmonic
representation for the closed cosmological models.
The super--Hamiltonian can be written as
\begin{equation}
H_{\perp}(x) = \sum_{(n)} H_{\perp}^{(n)} Q^{(n)}(x),
\label{H-perp-expansion}
\end{equation}
where $Q^{(n)}(x)$ is a scalar--type harmonic of the corresponding Laplace
operator, labeled by a multi-index $(n)$. The discrete set of constraints
$H_{\perp}^{(n)}$ is given by
\begin{equation}
H_{\perp}^{(n)} = \int dV  \, H_{\perp}(x)  Q^{*(n)}(x),
\label{H-perp-n}
\end{equation}
where $dV$ denotes the covariant volume element of the manifold; the harmonics
are normalized
\begin{equation}
 \int dV \, Q^{(n)}(x)  Q^{*(n)}(x) = \delta^{(n)(m)},
\label{scal-harm-norm}
\end{equation}
The structure of supermomentum expansion is more complicated:
\begin{equation}
H_i (x) = \sum_{(n)} H_{||}^{(n)} {\partial}_i Q^{(n)}(x) +
\sum_{(A)} H^{(A)} S_i^{(A)} (x) \, ,
\label{H-i-expansion}
\end{equation}
here $S_i^{(A)} (x)$ stand for the transverse vector harmonics obeying
the relations
\begin{equation}
{\nabla}^i S_i^{(A)} (x) = 0 \, .
\end{equation}
So, two types supermomentum constraint harmonics appear: the longitudinal
$H_{||}$ and the transversal $H^{(A)}$ labeled by another set of
multi--indices
$(A)$. The inverse relations for $H_{||}$ and $H_{(A)}$ are
\begin{equation}
H_{||}^{(n)} = \int dV  \, H_{i}(x)
\frac{{\partial}^i Q^{*(n)}(x)}{\lambda(n)} \, ,
\label{H-||-n}
\end{equation}
\begin{equation}
H^{(A)} = \int dV \, H_{i}(x) S^{* i (A)}(x) \, ,
\label{H-A}
\end{equation}
where ${\lambda}^{(n)}$ is an eigenvalue  of the
Laplace operator corresponding
to the eigenfunction $Q^{(n)}(x)$.
Transversal vector harmonics are normalized
\begin{equation}
\int dV \, S_i^{(A)} S^{i \, (B)} = {\delta}^{(A)(B)}.
\label{S-i}
\end{equation}
Transversal constraint's
harmonics $H^{(A)}$ define corresponding continuous subset of
supermomentum constraints
\begin{equation}
\tilde{H_i} (x) =
\sum_{(A)} H^{(A)} S_i^{(A)} (x) \, .
\label{Area-preserv-expansion}
\end{equation}
Constraints $\tilde{H_i}(x)$ are obviously divergenceless
\begin{equation}
{\nabla}^i \tilde{H_i}(x) =0 \, ,
\label{divergenless-of-area-preserve}
\end{equation}
where the covariant derivative is defined with respect to the metric of the
maximally symmetric space, which we shall call as a background metrics in
what follows.

It  is easy to check that the Poisson bracket of $\tilde{H_i}(x)$
 among themselves gives $\tilde{H_i}(x)$ again
\begin{equation}
\{ \tilde{H_i}(x) \, , \, \tilde{H_j}(x^{\prime}) \} =
\tilde{H_j}(x) \frac{\partial}{\partial x^i}
\delta (x, x^{\prime})
- \tilde{H_i}(x^{\prime}) \frac{\partial}{\partial x^{\prime j}}
\delta (x, x^{\prime})
\label{area-preserv-algebra}
\end{equation}
Usually, it is suggestive to consider the contractions of constraints
to arbitrary "weight--functions". In the case of $\tilde{H_i}(x) \, ,$
these constraint functionals
\begin{equation}
\int dV \tilde{H_i}(x) \, f^i (x) \equiv  \tilde{H}(f) \, ,
\end{equation}
where $f^i (x)$ can be chosen divergenceless,
form the closed algebra with respect to
Poisson bracket
\begin{equation}
\{  \tilde{H}(f_1) \, , \tilde{H}(f_2) \} =  \tilde{H} ( [ f_1 , f_2]) \, ,
\quad  [ f_1 , f_2]^i = f^j_1 \frac{\partial}{\partial x^j} f^i_2 -
f^j_2 \frac{\partial}{\partial x^j} f^i_1 \, .
\label{area-preserv-algebra1}
\end{equation}
The Lie bracket of divergenceless vector fields $f_1 , f_2 $,
 apparently,
is a divergenceless too.
The commutation relations (\ref{area-preserv-algebra1})
are typical for generators of area--preserving diffeomorphisms
introduced in the paper \cite{Arnold} and being intensively studied
in relationship with $p$-brane theories \cite{area-preserving}.
However, the transformations generated by $\tilde{H}$ do not preserve
the volume of the physical space, i.e. $\{\tilde{H},g\} \neq 0$,
where $ g = det \, g_{ij}$ is a genuine (not a background) metric.
In principle,
one can introduce divergenceless projection of supermomentum
with respect to the genuine metric $g_{ij}$, and such a projection
does generate area-preserving transformations. However, these
projections of the constraints do not form a closed algebra with respect to
Poisson bracket. This disclosure is caused by the fact that the
projector depends upon the metric and so it is not invariant with
respect to a metric shift. So, the Poisson bracket of two area-preserving 
supermomentum constraints does not give an
area-preserving projection of a supermomentum constraint again.
For our purposes it is
essential to employ the constraint basis, with the closed
subalgebra of spatial transversal diffeomorphisms, and we adopt for this 
reason the basis
of longitudinal and transversal constraints. The latter we shall refer
to as area-preserving, because they form the closed subalgebra of
this type, although they employ a background metric
instead of a genuine one. The longitudinal part of supermomentum will
be combined with the super-Hamiltonian discrete constraints
(\ref{H-perp-n}) to construct Virasoro-like generators.

In continuous basis these generators read
\begin{equation}
H_{\perp}(x) \pm \frac{1}{\Delta^{1/2}} \nabla^{i}H_{i}(x),
\label{Virasoro-like}
\end{equation}
where the covariant derivative $\nabla^{i}$ and Laplace operator $\Delta$
are defined in respect to background metric. Similarly,
area-preserving diffeomorphisms in continuous parameterization are
\begin{equation}
\left(\delta_{j}^{i} - \nabla_{j}
\frac{1}{\Delta}\nabla^{i}\right)H_{i}.
\label{area-cont}
\end{equation}

Dynamical variables are also expanded into the appropriate set of
harmonics. For example, a scalar field expansion is
\begin{equation}
\varphi(t,x) = \sum_{(n)} \varphi^{(n)}(t)Q^{(n)}(x),
\label{scal-harm}
\end{equation}
where $Q^{(n)}$ is the same set of scalar harmonics as in the
expansion of $H_{\perp}(x)$ (\ref{H-perp-expansion}). The expansion
of metrics is more complicated:
\begin{eqnarray}
&&g_{ij}(t,x) = \sum_{(n)} a^{(n)}(t) g_{ij}^{(0)}(x)Q^{(n)}(x)
+\sum_{(n)}b^{(n)}(t)\left(\frac{\nabla_{i}\nabla_{j}}{\lambda^{(n)}}
-\frac{g_{ij}^{(0)}}{N}\right)Q^{(n)}(x)\nonumber \\
&&+\sum_{(A)} c^{(A)}(t)(\nabla_{i}S_{j}^{(A)}(x)
+\nabla_{j}S_{i}^{(A)}(x))
+\sum_{(\alpha)}d^{(\alpha)}(t)G_{ij}^{(\alpha)}(x),
\label{metric-expansion}
\end{eqnarray}
where $g_{ij}^{(0)}(x)$ is the background metric,
$G_{ij}^{(\alpha)}(x)$ is the set of transverse traceless tensor harmonics:
\begin{equation}
g^{ij(0)}G_{ij} = 0; \qquad \nabla^{i}G_{ij} = 0.
\end{equation}
Similar expansions can be written
for the rest of canonical variables.
For the case of $S^{3}$ this multipole expansion (\ref{metric-expansion})
was described in detail in \cite{Khalat}. 

The expansion for $g_{ij}$ (\ref{metric-expansion}) is involved
not only into super-Hamiltonian (\ref{superhamiltonian}) and
supermomenta (\ref{supermomentum}), but also into the structure
functions of constraint algebra via relation (\ref{involution1}),
thereby the multipole modes $a^{(n)}(t), b^{(n)}(t), c^{(A)}(t)$
and $d^{(\alpha)}(t)$  enter the involution coefficients.
However, in the framework of the perturbative consideration
adopted in this paper for the calculation of quantum corrections,
the first approximation can be obtained neglecting the multipoles from
the involution structure functions. This will be explained in more
detail in Sec. 5, where quantum BRST charge will be studied for the
case of Bianchi-$I$ cosmology.

Let us discuss the multipole independent part of the structure
functions. One may easily see that they are proportional to the
integrals of three harmonic functions of such types:
\begin{eqnarray}
&&\int dV Q^{(n)}(x)Q^{(m)}(x)Q^{*(l)}(x),\nonumber\\
&&\int dV Q^{(n)}(x)Q^{(m)}(x)S_{i}^{*(A)}(x),\nonumber\\
&&\int dV Q^{(n)}(x)S_{i}^{*(A)}(x)S_{j}^{*(B)}(x),\nonumber\\
&&\int dV S_{i}^{(A)}(x)S_{J}^{(B)}(x)S_{k}^{*(C)}(x).
\label{triplets}
\end{eqnarray}

The integrals (\ref{triplets}) can be expressed via bilinear
combinations of Clebsch-Gordan coefficients of the corresponding
symmetry group by making use of Wigner-Eckart theorem (see e.g.
\cite{Wigner-Eckart}).

If the expansion of constraints into the harmonics of dynamical
modes is considered, then the coefficients at bilinear combinations
of multipoles are expressed in terms of the similar integrals, which
in their turn can be reduced to the Clebsch-Gordan coefficients.
As a matter of fact, only quadratic combinations of harmonics in
the constraints may contribute to the constant part of the first
quantum correction to the constraint involution relations.
This contribution can usually be very instructive, e.g.: for the
case of string \cite{Hwang,Okhta}, membranes \cite{membranes}
or $W_{3}$-gravity \cite{Tierry} it is the
correction, which fixes the critical parameters.

From the viewpoint of obtaining the critical parameters for closed
cosmological models, the method is basically the same for any group,
but the simplest case of Bianchi-$I$ cosmology allows to avoid the
cumbersome expressions involving summing of bilinear combinations
of Clebsch-Gordan coefficients.

\section{Wick ordering  
and basis of integer vectors on $N$-torus}

In contrast to ordinary quantum mechanics, quantum theories with an
infinite number of degrees of freedom may be inequivalent for a
different choices of operator symbols. Of course, different symbols
of operators (different orderings) are formally linked by unitary
transformations, but an appearance of ultraviolet divergences
may destroy the unitarity of the transformation. For example,
Wick symbols of the Virasoro constraints in bosonic string theory
yield the nonvanishing finite central extension, whereas Weyl symbol
of the same constraints does not reveal this correction to the
commutation relations. So, the choice of ordering appears to be
not technical, but fundamental step, defining the theory.

The common wisdom is that the modes of oscillating behaviour should
be quantized by Wick ordering, while those, having non-oscillating
dynamics (usually, zero modes) are ordered by Weyl rule.

To provide an opportunity of Wick quantization of the part of the
constraints and Fock representation for the model we should introduce
some kind of split of the set of $N$-dimensional wave vectors on $N$-torus.
In the case of string the similar problem can be resolved in a
very simple way because there are one-dimensional vectors (i.e. integer
numbers) $n$. One merely considers positive integer numbers as the
numbers corresponding to the annihilation operators $a_{n}$ and 
creation operators $\bar{a}_{n}^{+}$ and negative integer numbers
as numbers corresponding to the creation operators $a_{n}^{+}$
and annihilation operators $\bar{a}_{n}$. In turn, ghost operators
corresponding to Virasoro constraints $L_{n}$ and $\bar{L}_{-n}$,
where $n$ is positive are treated as creation operators while the
ghost operators corresponding to constraints $L_{-n}$ and
$\bar{L}_{n}$ are treated as annihilation operators.

An attempt to consider Virasoro-like constraint subset
for multidimensional
cases  should imply some splitting of integer vectors
\begin{equation}
\vec{n} = (n_{1},n_{2},\cdots,n_{N})
\label{wavevector}
\end{equation}
into positive and negative ones. The plausible way to do it, from
our point of view, is suggested in the paper \cite{Figuerido-Ramos}.

Let us fix in some way the order of main axis of our torus.
Then, if the first wave number $n_{1}$ is positive, we shall call the
vector positive and if $n_{1}$ is negative we shall call the vector
negative. If $n_{1} = 0$ we should turn to the second integer number
$n_{2}$ and to call the vector $\vec{n}$ positive or negative
depending on sign of $n_{2}$. If both numbers $n_{1}$ and $n_{2}$
are equal to zero, the sign of vector $\vec{n}$ is determined by the
sign of $n_{3}$ and so on. One can easily see that all the vectors
$\vec{n}$  must belong to one of these two classes excepting the
zero-vector $(0,0,\cdots,0)$. Below we shall implement a notation
\begin{eqnarray}
&sgn(\vec{n})& =  1 \, , \;if \quad \vec{n} \;is\;positive\nonumber\\
&sgn(\vec{n})& = -1 \; , \, if \quad \vec{n} \; is \;negative.
\label{signature}
\end{eqnarray}

As a matter of fact we shall consider the integer vectors weighted
by radii of $N$-torus:
\begin{equation}
\vec{n} =\left(\frac{n_{1}}{R_{1}},\ldots,\frac{n_{N}}{R_{N}}\right).
\label{wavevector1}
\end{equation}

In what follows we shall basically use the so called  ``last element'' limit,
i.e. we shall reduce our calculations mainly to the case then only number
$n_{N}$ differs from zero. It will give us an opportunity to make all
the calculations explicitly. 

\section{Constraint algebra of Bianchi-$I$ model \newline
in harmonic representation}

In this section, we write down explicitly Virasoro-like
and area-preserving diffeomorphism constraints
for Bianchi-$I$ model in harmonic representation and study their
commutation relations.
\begin{equation}
L(\vec{n}) = \frac{1}{2}\left(H_{\perp}(\vec{n}) + \frac{sgn(\vec{n})
n^{i} H_{i} (\vec{n})}{|\vec{n}|}\right),
\label{L-define}
\end{equation}
\begin{equation}
\bar{L}(\vec{n}) = \frac{1}{2}\left(H_{\perp}(\vec{n}) -
\frac{sgn(\vec{n}) n^{i} H_{i}(\vec{n})}{|\vec{n}|}\right),
\label{L-bar-define} 
\end{equation}
where
\begin{equation}
H_{\perp}(\vec{n}) = \int d^{N}x H_{\perp}(x)\exp(i\vec{n}\vec{x}),
\label{H-perp-n-define}
\end{equation}
\begin{equation}
H_{i}(\vec{n}) = \int d^{N}x H_{i}(x)\exp(i\vec{n}\vec{x}),
\label{H-i-n-define}
\end{equation}

Here, positive and negative constraint harmonics $L(\vec{n}),
\bar{L}(\vec{n})$ and $L(-\vec{n}), \bar{L}(-\vec{n})$ are complex
conjugated:
\begin{equation}
L^{*}(\vec{n}) = L(-\vec{n}), \qquad
\bar{L}^{*}(\vec{n}) = \bar{L}(-\vec{n}).
\label{conjug}
\end{equation}

Let us write down the involution relations among these constraints,
omitting the gravity harmonic contributions from structure
coefficients:
\begin{eqnarray}
&&[L(\vec{n}),L(\vec{m})] = U_{L(\vec{n}) L(\vec{m})}^{L(\vec{n}+\vec{m})}
L(\vec{n}+\vec{m}) + U_{L(\vec{n}) L(\vec{m})}^{\bar{L}(\vec{n}+\vec{m})}
\bar{L}(\vec{n}+\vec{m})
+U_{L(\vec{n}) L(\vec{m})}^{\tilde{H}_{i}(\vec{n}+\vec{m})}
\tilde{H}_{i}(\vec{n}+\vec{m})\nonumber\\
&&= \frac{1}{4}\left(1+\frac{sgn(\vec{n})
sgn(\vec{m})\vec{m}\vec{n}}{|\vec{m}||\vec{n}|}\right)\nonumber \\
&&\times\left\{\left(sgn(\vec{n})|\vec{n}| - sgn(\vec{m})|\vec{m}|
+\frac{sgn(\vec{n}+\vec{m})(\vec{n}^{2}-\vec{m}^{2})}{|\vec{n}+
\vec{m}|}\right)L(\vec{n}+\vec{m})\right.\nonumber \\
&&+\left(sgn(\vec{n})|\vec{n}| - sgn(\vec{m})|\vec{m}|
-\frac{sgn(\vec{n}+\vec{m})(\vec{n}^{2}-\vec{m}^{2})}{|\vec{n}+
\vec{m}|}\right)\bar{L}(\vec{n}+\vec{m})\nonumber \\
&&+\left.\left(n_{i}-m_{i} 
-\frac{(\vec{n}^{2}-\vec{m}^{2})(n_{i}+m_{i})}{(\vec{n}+\vec{m})^{2}}
\right)H_{i}(\vec{n}+\vec{m})\right\},
\label{commutator1}
\end{eqnarray}
\begin{eqnarray}
&&[L(\vec{n}),\bar{L}(\vec{m})] =
U_{L(\vec{n}) \bar{L}(\vec{m})}^{L(\vec{n}+\vec{m})}
L(\vec{n}+\vec{m}) +
U_{L(\vec{n}) \bar{L}(\vec{m})}^{\bar{L}(\vec{n}+\vec{m})}
\bar{L}(\vec{n}+\vec{m})
+U_{L(\vec{n}) \bar{L}(\vec{m})}^{\tilde{H}_{i}(\vec{n}+\vec{m})}
\tilde{H}_{i}(\vec{n}+\vec{m})\nonumber\\
&&= \frac{1}{4}\left(1-\frac{sgn(\vec{n})
sgn(\vec{m})\vec{m}\vec{n}}{|\vec{m}||\vec{n}|}\right)\nonumber \\
&&\times\left\{\left(sgn(\vec{n})|\vec{n}| + sgn(\vec{m})|\vec{m}|
+\frac{sgn(\vec{n}+\vec{m})(\vec{n}^{2}-\vec{m}^{2})}{|\vec{n}+
\vec{m}|}\right)L(\vec{n}+\vec{m})\right.\nonumber \\
&&+\left(sgn(\vec{n})|\vec{n}| + sgn(\vec{m})|\vec{m}|
-\frac{sgn(\vec{n}+\vec{m})(\vec{n}^{2}-\vec{m}^{2})}{|\vec{n}+
\vec{m}|}\right)\bar{L}(\vec{n}+\vec{m})\nonumber \\
&&+\left.\left(n_{i}-m_{i} 
-\frac{(\vec{n}^{2}-\vec{m}^{2})(n_{i}+m_{i})}{(\vec{n}+\vec{m})^{2}}
\right)H_{i}(\vec{n}+\vec{m})\right\},
\label{commutator2}
\end{eqnarray}
\begin{eqnarray}
&&[\bar{L}(\vec{n}),\bar{L}(\vec{m})] =
U_{\bar{L}(\vec{n}) \bar{L}(\vec{m})}^{L(\vec{n}+\vec{m})}
L(\vec{n}+\vec{m}) +
U_{\bar{L}(\vec{n}) \bar{L}(\vec{m})}^{\bar{L}(\vec{n}+\vec{m})}
\bar{L}(\vec{n}+\vec{m})
+U_{\bar{L}(\vec{n}) \bar{L}(\vec{m})}^{\tilde{H}_{i}(\vec{n}+\vec{m})}
\tilde{H}_{i}(\vec{n}+\vec{m})\nonumber\\
&&= \frac{1}{4}\left(1+\frac{sgn(\vec{n})
sgn(\vec{m})\vec{m}\vec{n}}{|\vec{m}||\vec{n}|}\right)\nonumber \\
&&\times\left\{\left(sgn(\vec{m})|\vec{m}| - sgn(\vec{n})|\vec{n}|
+\frac{sgn(\vec{n}+\vec{m})(\vec{n}^{2}-\vec{m}^{2})}{|\vec{n}+
\vec{m}|}\right)L(\vec{n}+\vec{m})\right.\nonumber \\
&&+\left(sgn(\vec{m})|\vec{m}| + sgn(\vec{n})|\vec{n}|
-\frac{sgn(\vec{n}+\vec{m})(\vec{n}^{2}-\vec{m}^{2})}{|\vec{n}+
\vec{m}|}\right)\bar{L}(\vec{n}+\vec{m})\nonumber \\
&&+\left.\left(n_{i}-m_{i} 
-\frac{(\vec{n}^{2}-\vec{m}^{2})(n_{i}+m_{i})}{(\vec{n}+\vec{m})^{2}}
\right)H_{i}(\vec{n}+\vec{m})\right\}.
\label{commutator3}
\end{eqnarray}

In the right-hand side of the relations (\ref{commutator1})--
(\ref{commutator3}) appear supermomenta constraints
$H_{i}(\vec{n}+\vec{m})$ contracted with vector
$$
\vec{n} - \vec{m} - \frac{(\vec{n}^{2}-\vec{m}^{2}) (\vec{n}+\vec{m})}
{(\vec{n}+\vec{m})^{2}},
$$
which is orthogonal to $(\vec{n}+\vec{m})$. Thus, we have seen, that
in the involution relations for Virasoro-like constraints $\L(\vec{n})$
and $\bar{L}(\vec{n})$ appear area-preserving diffeomorphisms.
So, the constraints $L, \bar{L}$ do not form a closed algebra.

It is convenient to describe area-preserving diffeomorphisms as
\begin{equation}
\tilde{H}_{\vec{v}}(\vec{n}) = v_{i}
H_{i}(\vec{n}),\; \qquad \vec{v} \vec{n} = 0.
\label{area-def}
\end{equation}
introducing an arbitrary vector $\vec{v}$, being orthogonal to the integer
vector $\vec{n}$.
The other involution relations are as follows:
\begin{eqnarray}
&&[\tilde{H}_{\vec{v}}(\vec{n}),L(\vec{m})] =
U_{\tilde{H}(\vec{v},\vec{n}) L(\vec{m})}^{L(\vec{n}+\vec{m})}
L(\vec{n}+\vec{m}) +
U_{\tilde{H}(\vec{v},\vec{n}) L(\vec{m})}^{\bar{L}(\vec{n}+\vec{m})}
\bar{L}(\vec{n}+\vec{m})
+U_{\tilde{H}(\vec{v},\vec{n})
L(\vec{m})}^{\tilde{H}(\vec{w},\vec{n}+\vec{m})}
\tilde{H}_{\vec{w}}(\vec{n}+\vec{m})
\nonumber \\
&&= - \frac{1}{2}(\vec{m}\vec{v})\left(1
+\frac{sgn(\vec{m})sgn(\vec{n}+\vec{m})|\vec{m}|}{|\vec{m}+\vec{n}|}
\right)L(\vec{n}+\vec{m})\nonumber \\
&&-\frac{1}{2}(\vec{m}\vec{v})\left(1
-\frac{sgn(\vec{m})sgn(\vec{n}+\vec{m})|\vec{m}|}{|\vec{m}+\vec{n}|}
\right)\bar{L}(\vec{n}+\vec{m})\nonumber \\
&&+\frac{1}{2}\frac{sgn(\vec{m})}{|\vec{m}|}
\left((\vec{m}\vec{n})v_{i}-(\vec{m}\vec{v})m_{i}
+\frac{\vec{m}^{2}(\vec{m}\vec{v})(n_{i}+m_{i})}{(\vec{n}+\vec{m})^{2}}
\right)H_{i}(\vec{n}+\vec{m}),
\label{commutator4}
\end{eqnarray}
\begin{eqnarray}
&&[H_{\vec{v}}(\vec{n}),\bar{L}(\vec{m})] =
U_{\tilde{H}(\vec{v},\vec{n}) \bar{L}(\vec{m})}^{L(\vec{n}+\vec{m})}
L(\vec{n}+\vec{m}) +
U_{\tilde{H}(\vec{v},\vec{n}) \bar{L}(\vec{m})}^{\bar{L}(\vec{n}+\vec{m})}
\bar{L}(\vec{n}+\vec{m})
+U_{\tilde{H}(\vec{v},\vec{n})
\bar{L}(\vec{m})}^{\tilde{H}(\vec{w},\vec{n}+\vec{m})}
\tilde{H}_{\vec{w}}(\vec{n}+\vec{m})
\nonumber \\
&&= - \frac{1}{2}(\vec{m}\vec{v})\left(1
-\frac{sgn(\vec{m})sgn(\vec{n}+\vec{m})|\vec{m}|}{|\vec{m}+\vec{n}|}
\right)L(\vec{n}+\vec{m})\nonumber \\
&&-\frac{1}{2}(\vec{m}\vec{v})\left(1
-\frac{sgn(\vec{m})sgn(\vec{n}+\vec{m})|\vec{m}|}{|\vec{m}+\vec{n}|}
\right)\bar{L}(\vec{n}+\vec{m})\nonumber \\
&&+\frac{1}{2}\frac{sgn(\vec{m})}{|\vec{m}|}
\left((\vec{m}\vec{n})v_{i}-(\vec{m}\vec{v})m_{i}
-\frac{\vec{m}^{2}(\vec{m}\vec{v})(n_{i}+m_{i})}{(\vec{n}+\vec{m})^{2}}
\right)H_{i}(\vec{n}+\vec{m}),
\label{commutator5}
\end{eqnarray}
\begin{eqnarray}
&&[H_{\vec{v}}(\vec{n}),H_{\vec{w}}(\vec{m})] =
U_{\tilde{H}(\vec{v},vec{n}) \tilde{H}(\vec{w},vec{m})}
^{\tilde{H}(\vec{u},\vec{n}+\vec{m})}\tilde{H}_{\vec{u}}(\vec{n}+\vec{m})
=
\nonumber \\
&&(\vec{m}\vec{w}) H_{\vec{v}}(\vec{n}+\vec{m})
-(\vec{n}\vec{v}) H_{\vec{w}}(\vec{n}+\vec{m}).
\label{commutator6}
\end{eqnarray}

Concluding this section we should notice that the formulae
(\ref{L-define}) and (\ref{L-bar-define})contain an ambiguity concerning
zero modes
$L(0)$ and $\bar{L}(0)$ which have to be described additionally. It
is convenient define them as
\begin{equation}
L(0) = \frac{1}{2}(H_{\perp}(0) + H_{N}(0))
\label{L-zero}
\end{equation}
and
\begin{equation}
\bar{L}(0) = \frac{1}{2}(H_{\perp}(0) - H_{N}(0))
\label{L-zero1}
\end{equation}
to provide a proper correspondence with a genuine Virasoro algebra in the
limiting case $N=1$.

\section{Quantum corrections to the algebra \newline
of Virasoro-like constraints}

In Sec. 3 we have introduced the split of integer vectors on $N$-torus
into positive and negative ones. Now we shall apply this decomposition
to define creation and annihilation operators for gravity and matter
fields, and to split the Virasoro-like constraints into positive and
negative ones. The latter decomposition will further define the natural
Wick ordering for the ghost variables upon the construction of quantum
BRST charge. Then we calculate the $c$-number part of the quantum
correction to the commutator $[L(\vec{n}),L(-\vec{n})]$, which is of
crucial importance from the standpoint of the nilpotency condition
on BRST operator.

To begin with, we consider the massless scalar field contributions
to constraints $H_{\perp}(\vec{n})$ and $H_{i}(\vec{n})$ which, in turn,
define Virasoro-like constraints. Scalar contributions into
super-Hamiltonian and supermomentum are as follows:
\begin{equation}
H_{\perp \;scalar} = \frac{1}{2} p^{2} 
+ \frac{1}{2} g g^{ij}\varphi,_{i}\varphi,_{j},
\label{superhamiltonian-scal}
\end{equation}
\begin{equation}
H_{i\;scalar} = p \varphi,_{i}.
\label{supermomentum-scal}
\end{equation}
Now, let us expand $\varphi$ and its conjugate momentum $p$ via
creation and annihilation operators:
\begin{eqnarray} 
&&\varphi = 
\frac{\varphi_{0}}{\sqrt{V}} + \sum_{\vec{k}>0} 
\frac{1}{\sqrt{2\omega_{\vec{k}}}\sqrt{V}}
(a_{\vec{k}}\exp(-i\vec{k}\vec{x})
+\bar{a}_{\vec{k}}\exp(i\vec{k}\vec{x})\nonumber \\
&&+a_{\vec{k}}^{+}\exp(i\vec{k}\vec{x})
+\bar{a}_{\vec{k}}^{+}\exp(-i\vec{k}\vec{x})),
\label{varphi-via-a}
\end{eqnarray}
\begin{eqnarray}
&&p = \frac{p_{0}}{\sqrt{V}} + i\sum_{\vec{k}>0} 
\sqrt{\frac{\omega_{\vec{k}}}{2V}}(-a_{\vec{k}}\exp(-i\vec{k}\vec{x})
-\bar{a}_{\vec{k}}\exp(i\vec{k}\vec{x})\nonumber \\
&&+a_{\vec{k}}^{+}\exp(i\vec{k}\vec{x})
+\bar{a}_{\vec{k}}^{+}\exp(-i\vec{k}\vec{x})).
\label{p-via-a}
\end{eqnarray}

Here $V$ is the volume of $N$-torus:
\begin{equation}
V = (2\pi)^{N} R_{1}R_{2} \cdots R_{N},
\label{volume-torus}
\end{equation}
where, $R_{i}$ are radii of the torus, $\omega_{\vec{k}} =
\sqrt{\vec{k}^{2}}$

Substituting Eqs. (\ref{varphi-via-a})--(\ref{p-via-a}) into
Eqs. (\ref{superhamiltonian-scal})--(\ref{supermomentum-scal})
and keeping only quadratic contributions one
can write down the following expressions for $H_{\perp \;scalar}
(\vec{n})$ and $H_{i \;scalar}(\vec{n})$
\begin{eqnarray}
&&H_{\perp \;scalar\; \vec{n}} = i p_{0} (\bar{a}_{\vec{n}}^{+}
-a_{\vec{n}})\sqrt{\frac{\omega_{\vec{k}}}{2}}\nonumber \\
&&-\frac{1}{2}\sum_{0<\vec{k}<\vec{n}}(a_{\vec{k}}a_{\vec{n}-\vec{k}}
+\bar{a}_{\vec{k}}^{+}\bar{a}_{\vec{n}-\vec{k}}^{+})
\left(\frac{\sqrt{\omega_{\vec{k}}\omega_{\vec{n}-\vec{k}}}}{2}
+\frac{\vec{k}(\vec{n}-\vec{k})}{2\sqrt{\omega_{\vec{k}}
\omega_{\vec{n}-\vec{k}}}}\right)\nonumber \\
&&+\frac{1}{2}\sum_{\vec{k}>0}(a_{\vec{n}+\vec{k}}\bar{a}_{\vec{k}}
+a_{\vec{k}}^{+}\bar{a}_{\vec{n}+\vec{k}}^{+})
\left(\frac{(\vec{n}+\vec{k})\vec{k}}
{\sqrt{\omega_{\vec{n}+\vec{k}}\omega_{\vec{k}}}}
-\sqrt{\omega_{\vec{n}+\vec{k}}\omega_{\vec{k}}}\right)\nonumber \\
&&+\frac{1}{2}\sum_{\vec{k}>0}(a_{\vec{n}+\vec{k}}a_{\vec{k}}^{+}
+\bar{a}_{\vec{k}}\bar{a}_{\vec{n}+\vec{k}}^{+})
\left(\sqrt{\omega_{\vec{n}+\vec{k}}\omega_{\vec{k}}}
+\frac{(\vec{n}+\vec{k})\vec{k}}
{\sqrt{\omega_{\vec{n}+\vec{k}}\omega_{\vec{k}}}}\right),
\label{superhamiltonian-scal1}
\end{eqnarray}
\begin{eqnarray}
&&H_{i\;scalar\;\vec{n}} = 
-ip_{0}\frac{n_{i}}{\sqrt{2\omega_{\vec{n}}}}(a_{\vec{n}}
+\bar{a}_{\vec{n}}^{+})\nonumber \\
&&+\sum_{0<\vec{k}<\vec{n}}(\bar{a}_{\vec{k}}^{+}\bar{a}_{\vec{n}-
\vec{k}}^{+}-a_{\vec{k}}a_{\vec{n}-\vec{k}})\frac{(n-k)_{i}}{2}
\sqrt{\frac{\omega_{\vec{k}}}{\omega_{\vec{n}-\vec{k}}}}\nonumber \\
&&+\sum_{\vec{k}>0}(a_{\vec{n}+\vec{k}}\bar{a}_{\vec{k}}
+a_{\vec{n}+\vec{k}}a_{\vec{k}}^{+})
\frac{k_{i}}{2} 
\sqrt{\frac{\omega_{\vec{n}+\vec{k}}}{\omega_{\vec{k}}}}\nonumber \\
&&-\sum_{\vec{k}>0}(\bar{a}_{\vec{k}}a_{\vec{n}+\vec{k}}+
\bar{a}_{\vec{k}}\bar{a}_{\vec{n}+\vec{k}}^{+})\frac{(n+k)_{i}}{2}
\sqrt{\frac{\omega_{\vec{k}}}{\omega_{\vec{n}+\vec{k}}}}\nonumber \\
&&+\sum_{\vec{k}>0}(a_{\vec{k}}^{+}a_{\vec{n}+\vec{k}}+
a_{\vec{k}}^{+}\bar{a}_{\vec{n}+\vec{k}}^{+})\frac{(n+k)_{i}}{2}
\sqrt{\frac{\omega_{\vec{k}}}{\omega_{\vec{n}+\vec{k}}}}\nonumber \\ 
&&-\sum_{\vec{k}>0}(\bar{a}_{\vec{n}+\vec{k}}^{+}\bar{a}_{\vec{k}}
+\bar{a}_{\vec{n}+\vec{k}}^{+}a_{\vec{k}}^{+})
\frac{k_{i}}{2} 
\sqrt{\frac{\omega_{\vec{n}+\vec{k}}}{\omega_{\vec{k}}}}.
\label{supermomentum-scal1}
\end{eqnarray}
The dynamical metric does not contribute into the expansions
(\ref{superhamiltonian-scal1}) and (\ref{supermomentum-scal1})
in quadratic approximation because the background value of
scalar field is zero. As a matter of fact, such a choice
of the background is the only
consistent for the stationary Bianchi-$I$ universe at a classical level.

It is easy to see from Eq. (\ref{superhamiltonian-scal1}) that
in the case when $\vec{n} = 0$, Hamiltonian takes the diagonal form
\begin{equation}
H_{\perp \;scalar\;\vec{0}} =  \frac{p_{0}^{2}}{2}
+\frac{1}{2}\sum_{\vec{k}>0}\omega_{\vec{k}}
(a_{\vec{k}}^{+}a_{\vec{k}} + 
\bar{a}_{\vec{k}}^{+}\bar{a}_{\vec{k}}).
\label{superhamiltonian-scal2}
\end{equation}
The choice of creation and
annihilation operators, given by Eqs. (\ref{varphi-via-a}) and
(\ref{p-via-a}), proves itself in the suggestive form of the
super-Hamiltonian.

Using formulae (\ref{superhamiltonian-scal1})--
(\ref{supermomentum-scal1}) one can get the expression for the 
contribution of the scalar fields into the central extension
(we mean here that zero modes for scalar field and its conjugate
momentum, as well as zero modes for other fields, are ordered according
to Weyl rule, while other modes are Wick-ordered)
\begin{eqnarray}
&&[L(\vec{n}),L(-\vec{n})]_{c.e.\;scalars} 
=\frac{1}{2}\frac{n_{i}}{|\vec{n}|} [H_{\perp}(\vec{n}),H_{i}(-\vec{n})]
_{c.e.scalars}\nonumber\\
&&\frac{n_{i}}{|\vec{n}|}
\left\{\frac{1}{4}\sum_{0<\vec{k}<\vec{n}}\left(\omega_{\vec{k}}(n-k)_{i}
+\frac{\vec{k}(\vec{n}-\vec{k})(n-k)_{i}}{\omega_{\vec{n}-\vec{k}}}\right)
\right.\nonumber \\
&&\left.+\frac{1}{8}\sum_{\vec{k}>0}\left(
\frac{(\vec{n}+\vec{k})\vec{k}}
{\sqrt{\omega_{\vec{n}+\vec{k}}\omega_{\vec{k}}}}
-\sqrt{\omega_{\vec{n}+\vec{k}}\omega_{\vec{k}}}\right)
\left(k_{i}\sqrt{\frac{\omega_{\vec{n}+\vec{k}}}{\omega_{\vec{k}}}}
-(n+k)_{i}\sqrt{\frac{\omega_{\vec{k}}}{\omega_{\vec{n}+\vec{k}}}}\right)
\right\}.
\label{central-scal}
\end{eqnarray}

In the ``last-element'' limit (see sec. 3)
the formula (\ref{central-scal}) is
reduced to 
\begin{equation}
[L(\vec{n}),L(-\vec{n})]_{c.e.\;scalars} = 
\frac{1}{2}\sum_{k=1}^{n-1} k(n-k) = \frac{1}{12}n(n^{2}-1),
\label{central-scal1}
\end{equation} 
and one can see that the contribution of scalar fields has 
a string-like structure.
In the case when the model includes $d$ massless scalar fields each of
them gives the same contribution to the commutator $[L(\vec{n}),
L(-\vec{n})]$, and thereby the right-hand side of (\ref{central-scal1})
is multiplied by $d$.

Consider now Wick representation for the metric and the quadratic part
of gravity constraints. It is convenient to rewrite quadratic contribution
of gravity variables into the constraints (\ref{superhamiltonian}) and
(\ref{supermomentum}) in the following form:

\begin{equation}
H_{i} = g_{ab} \pi^{cd},_{e} E_{i}^{ab},_{cd},^{e}
+ g_{ab,e} \pi^{cd} F_{i}^{ab},_{cd},^{e}
\label{supermomentum1}
\end{equation}
and
\begin{equation}
H_{\perp} = l_{P}^{N-1} G_{ab,cd}\pi^{ab}\pi^{cd} + 
\frac{1}{l_{P}^{N-1}} C^{ab,cd,ef}g_{ab}g_{cd,ef} 
+\frac{1}{l_{P}^{N-1}} D^{ab,e,cd,f}g_{ab,e}g_{cd,f}.
\label{superhamiltonian1}
\end{equation}
Here, in the Bianchi-I flat background, we have:
\begin{equation}
E_{i}^{ab},_{cd},^{e} = -2 \delta 
^{(a}_{(c}\delta^{e}_{d)}\delta^{b)}_{i},
\label{E-define}
\end{equation}
\begin{equation}
F_{i}^{ab},_{cd},^{e} = \delta^{ab}_{cd} \delta^{e}_{i} 
-2 \delta^{ab}_{i(c} \delta^{e}_{d)},
\label{F-define}
\end{equation}
where
\begin{equation}
\delta^{ab}_{cd} \equiv \delta^{(a}_{(c} \delta^{b)}_{d)} \equiv \frac{1}{2}
({\delta}^{a}_c {\delta}^{b}_d + {\delta}^{b}_c {\delta}^{a}_d) \, .
\label{delta-define}
\end{equation}
\begin{eqnarray}
&&C^{ab,cd,ef} = \delta^{ab}\delta^{cd}\delta_{ef} -
\delta^{ab} \delta^{(c(e}\delta^{d)f}
-\delta^{(a(c}\delta^{b)d)}\delta^{ef}\nonumber \\ 
&&-\delta^{(a(e}\delta^{b)f)}\delta^{cd}
+2\delta^{(a(c}\delta^{b)(e}\delta^{d)f)}.
\label{C-define}
\end{eqnarray}
\begin{eqnarray}
&&D^{ab,e,cd,f} = \frac{1}{4}\delta^{ab}\delta^{cd}\delta_{ef}
-\frac{1}{2}\delta^{ab} \delta^{(c(e}\delta^{d)f)}
-\frac{1}{2} \delta^{(a(e}\delta^{b)f)}\delta^{cd}\nonumber\\
&&-\frac{3}{4}\delta^{(a(c}\delta^{b)d)}\delta^{ef}
+\delta^{(a(c}\delta^{b)(e}\delta^{d)f)}
+\frac{1}{2}\delta^{(af}\delta^{b)(d}\delta^{c)e}.
\label{D-define}
\end{eqnarray}
In Eqs. (\ref{E-define})--(\ref{D-define}) we have symmetrization
in couples of indices $a$ and $b$, $c$ and $d$ and in Eq. 
(\ref{C-define}) we have also the symmetrization in indices $e$ and 
$f$.

Now let us expand the metric $g_{ab}$ and
the conjugate momentum $\pi^{ab}$ in terms of
operators of creation and annihilation corresponding to different 
harmonics:  
\begin{eqnarray} 
&&g_{ab}(\vec{x}) = \sqrt{\frac{2l_{P}^{N-1}}{V}}g_{ab}(0) + 
\sum_{\vec{k}} 
\sqrt{\frac{2l_{P}^{N-1}}{V}}\sqrt{\frac{1}{2\omega_{\vec{k}}}} 
(a_{ab}(\vec{k})\exp(-i\vec{k}\vec{x}) \nonumber\\
&&+a_{ab}^{+}(\vec{k})\exp(i\vec{k}\vec{x})
+\bar{a}_{ab}(\vec{k})\exp(i\vec{k}\vec{x})
+\bar{a}_{ab}^{+}(\vec{k})\exp(-i\vec{k}\vec{x})),
\label{g-via-a}
\end{eqnarray}
\begin{eqnarray} 
&&\pi^{cd}(\vec{x}) =  \sqrt{\frac{1}{2l_{P}^{N-1}V}}\pi^{cd}(0) + 
\sum_{\vec{k}} i G^{ab,cd} 
\sqrt{\frac{1}{2l_{P}^{N-1}V}}
\sqrt{\frac{\omega_{\vec{k}}}{2}} 
(-a_{ab}(\vec{k})\exp(-i\vec{k}\vec{x}) \nonumber\\
&&+a_{ab}^{+}(\vec{k})\exp(i\vec{k}\vec{x})
-\bar{a}_{ab}(\vec{k})\exp(i\vec{k}\vec{x})
+\bar{a}_{ab}^{+}(\vec{k})\exp(-i\vec{k}\vec{x})).
\label{pi-via-a}
\end{eqnarray}
It is the choice of creation and annihilation operators which provides
the diagonal form for the part of Hamiltonian $H_{\perp
}(\vec{0})$ describing physical degrees of freedom for gravitons, i.e.
transverse-traceless modes.

Now one can write down the $n$-th component of super-Hamiltonian
(\ref{H-perp-n-define}) and supermomentum (\ref{H-i-n-define})
in terms of expansions (\ref{g-via-a}) and (\ref{pi-via-a}):

\begin{eqnarray}
&&H_{\perp}(\vec{n}) = \frac{i}{2}\sqrt{2\omega_{\vec{n}}}\pi^{ab}(0)
(-a_{ab}(\vec{n}) + \bar{a}_{ab}^{+}(\vec{n}))\nonumber \\
&&-\frac{2n_{e}n_{f}}{\sqrt{2\omega_{\vec{n}}}}
C^{ab,cd,ef}g_{ab}(0)(a_{cd}(\vec{n}) +
\bar{a}_{ab}^{+}(\vec{n}))\nonumber \\
&&-\sum_{0<\vec{k}<\vec{n}}(a_{ab}(\vec{k})a_{cd}(\vec{n}-\vec{k})
+\bar{a}_{ab}^{+}(\vec{k})\bar{a}_{cd}^{+}(\vec{n}-\vec{k}))\nonumber 
\\ 
&&\times\left(\frac{\sqrt{\omega_{\vec{k}}\omega_{\vec{n}-\vec{k}}}}{4} 
G^{ab,cd} + \frac{(n-k)_{e}(n-k)_{f}} 
{2\sqrt{\omega_{\vec{k}}\omega_{\vec{n}-\vec{k}}}}C^{ab,cd,ef}
+\frac{k_{e}k_{f}}
{2\sqrt{\omega_{\vec{k}}\omega_{\vec{n}-\vec{k}}}}C^{cd,ab,ef}\right.
\nonumber\\
&&\left.+\frac{k_{e}(n-k)_{f}}
{2\sqrt{\omega_{\vec{k}}\omega_{\vec{n}-\vec{k}}}}D^{ab,e,cd,f}
  +\frac{(n-k)_{e}k_{f}}
{2\sqrt{\omega_{\vec{k}}\omega_{\vec{n}-\vec{k}}}}D^{cd,f,ab,e}\right)
\nonumber \\
&&+\sum_{0<\vec{k}<\vec{n}}(a_{ab}(\vec{k})
\bar{a}_{cd}^{+}(\vec{n}-\vec{k})
+\bar{a}_{ab}^{+}(\vec{k})a_{cd}(\vec{n}-\vec{k}))\nonumber \\
&&\times\left(\frac{\sqrt{\omega_{\vec{k}}\omega_{\vec{n}-\vec{k}}}}{4}
G^{ab,cd}-\frac{(n-k)_{e}(n-k)_{f}}
{\sqrt{\omega_{\vec{k}}\omega_{\vec{n}-\vec{k}}}}C^{ab,cd,ef}
-\frac{k_{e}(n-k)_{f}}
{\sqrt{\omega_{\vec{k}}\omega_{\vec{n}-\vec{k}}}}
D^{ab,ecd,f}\right)\nonumber \\
&&+\sum_{\vec{k}>0} (a_{ab}(\vec{n}+\vec{k})a_{cd}^{+}(\vec{k})
+\bar{a}_{ab}^{+}(\vec{n}+\vec{k})\bar{a}_{cd}(\vec{k}))\nonumber \\
&&\times 
\left(\frac{\sqrt{\omega_{\vec{k}}\omega_{\vec{n}+\vec{k}}}}{4} 
G^{ab,cd}-\frac{k_{e}k_{f}}
{\sqrt{\omega_{\vec{k}}\omega_{\vec{n}+\vec{k}}}}C^{ab,cd,ef}
+\frac{(n+k)_{e}k_{f}}
{\sqrt{\omega_{\vec{k}}\omega_{\vec{n}+\vec{k}}}}
D^{ab,ecd,f}\right)\nonumber \\
&&+\sum_{\vec{k}>0} (a_{ab}(\vec{n}+\vec{k})\bar{a}_{cd}(\vec{k})
+\bar{a}_{ab}^{+}(\vec{n}+\vec{k})a_{cd}^{+}(\vec{k}))\nonumber \\
&&\times 
\left(-\frac{\sqrt{\omega_{\vec{k}}\omega_{\vec{n}+\vec{k}}}}{4} 
G^{ab,cd}-\frac{k_{e}k_{f}}
{\sqrt{\omega_{\vec{k}}\omega_{\vec{n}+\vec{k}}}}C^{ab,cd,ef}
+\frac{(n+k)_{e}k_{f}}
{\sqrt{\omega_{\vec{k}}\omega_{\vec{n}+\vec{k}}}}
D^{ab,ecd,f}\right)\nonumber \\
&&+\sum_{\vec{k}>0} (a_{ab}^{+}(\vec{k})a_{cd}(\vec{n}+\vec{k})
+\bar{a}_{ab}(\vec{k})a_{cd}^{+}(\vec{n}+\vec{k}))\nonumber \\
&&\times 
\left(\frac{\sqrt{\omega_{\vec{k}}\omega_{\vec{n}+\vec{k}}}}{4} 
G^{ab,cd}-\frac{(n+k)_{e}(n+k)_{f}}
{\sqrt{\omega_{\vec{k}}\omega_{\vec{n}+\vec{k}}}}C^{ab,cd,ef}
+\frac{k_{e}(n+k)_{f}}
{\sqrt{\omega_{\vec{k}}\omega_{\vec{n}+\vec{k}}}}
D^{ab,ecd,f}\right)\nonumber \\
&&+\sum_{\vec{k}>0} (\bar{a}_{ab}(\vec{k})a_{cd}(\vec{n}+\vec{k})
+a_{ab}^{+}(\vec{k})\bar{a}_{cd}^{+}(\vec{n}+\vec{k}))\nonumber \\
&&\times 
\left(-\frac{\sqrt{\omega_{\vec{k}}\omega_{\vec{n}+\vec{k}}}}{4} 
G^{ab,cd}-\frac{(n+k)_{e}(n+k)_{f}}
{\sqrt{\omega_{\vec{k}}\omega_{\vec{n}+\vec{k}}}}C^{ab,cd,ef}
+\frac{k_{e}(n+k)_{f}}
{\sqrt{\omega_{\vec{k}}\omega_{\vec{n}+\vec{k}}}}
D^{ab,ecd,f}\right),
\label{superhamiltonian-n1}
\end{eqnarray}
\begin{eqnarray}
&&H_{i \vec{n}} = g_{ab}(0)(\bar{a}_{cd}^{+}(\vec{n})-a_{cd}(\vec{n}))
n_{e}\sqrt{\frac{\omega_{\vec{n}}}{2}}E_{i}^{ab,cd,e}\nonumber \\
&&-i\pi_{cd}(0)(a_{ab}(\vec{n})+\bar{a}_{ab}^{+}(\vec{n}))
\frac{n_{e}}{\sqrt{2\omega_{\vec{n}}}}F_{i}^{ab,cd,e}\nonumber \\
&&+\sum_{0<\vec{k}<\vec{n}} (\bar{a}_{ab}^{+}(\vec{k})\bar{a}_{cd}^{+}
(\vec{n}-\vec{k})-a_{ab}(\vec{k})a_{cd}(\vec{n}-\vec{k}))\nonumber \\
&&\times\left(\frac{(n-k)_{e}}{4}\sqrt{\frac{\omega_{\vec{n}-\vec{k}}}
{\omega_{\vec{k}}}}E_{i}^{ab,cd,e} + 
\frac{k_{e}}{4}\sqrt{\frac{\omega_{\vec{k}}}{\omega_{\vec{n}-\vec{k}}}}
E_{i}^{ab,cd,e}\right.\nonumber \\
&&\left.+\frac{k_{e}}{4}\sqrt{\frac{\omega_{\vec{n}-\vec{k}}}
{\omega_{\vec{k}}}}F_{i}^{ab,cd,e} + 
\frac{(n-k)_{e}}{4}\sqrt{\frac{\omega_{\vec{k}}}
{\omega_{\vec{n}-\vec{k}}}}
F_{i}^{ab,cd,e}\right)\nonumber \\
&&+\sum_{0<\vec{k}<\vec{n}}
(a_{ab}(\vec{k})\bar{a}_{cd}^{+}
(\vec{n}-\vec{k})-\bar{a}_{ab}^{+}(\vec{k})a_{cd}(\vec{n}-\vec{k}))
\nonumber \\
&&\times\left(\frac{(n-k)_{e}}{2}\sqrt{\frac{\omega_{\vec{n}-\vec{k}}}
{\omega_{\vec{k}}}}E_{i}^{ab,cd,e} + 
\frac{k_{e}}{2}\sqrt{\frac{\omega_{\vec{n}-\vec{k}}}
{\omega_{\vec{k}}}}F_{i}^{ab,cd,e}\right)\nonumber \\
&&+\sum_{\vec{k}>0}(a_{ab}(\vec{n}+\vec{k})\bar{a}_{cd}(\vec{k})
-\bar{a}_{ab}^{+}(\vec{n}+\vec{k})a_{cd}^{+}(\vec{k})\nonumber \\
&&-a_{ab}(\vec{n}+\vec{k})a_{cd}^{+}(\vec{k})
+\bar{a}_{ab}^{+}(\vec{n}+\vec{k})\bar{a}_{cd}(\vec{k}))
\nonumber \\
&&\times\left(\frac{k_{e}}{2}
\sqrt{\frac{\omega_{\vec{k}}}{\omega_{\vec{n}+\vec{k}}}}
E_{i}^{ab,cd,e}-\frac{(n+k)_{e}}{2}
\sqrt{\frac{\omega_{\vec{k}}}{\omega_{\vec{n}+\vec{k}}}}
F_{i}^{ab,cd,e}\right)\nonumber \\
&&+\sum_{\vec{k}>0}(a_{ab}^{+}(\vec{k})\bar{a}_{cd}^{+}(\vec{n}+\vec{k})
-a_{ab}^{+}(\vec{k})a_{cd}(\vec{n}+\vec{k})\nonumber \\
&&-\bar{a}_{ab}(\vec{k})a_{cd}(\vec{n}+\vec{k})
+\bar{a}_{ab}(\vec{k})\bar{a}_{cd}^{+}(\vec{n}+\vec{k}))
\nonumber \\
&&\times\left(\frac{(n+k)_{e}}{2}\sqrt{\frac{\omega_{\vec{n}+\vec{k}}}
{\omega_{\vec{k}}}}E_{i}^{ab,cd,e} -
\frac{k_{e}}{2}\sqrt{\frac{\omega_{\vec{n}+\vec{k}}}
{\omega_{\vec{k}}}}F_{i}^{ab,cd,e}\right).
\label{supermomentum-n1}
\end{eqnarray}

Let us mention that the approximation of constraints by quadratic
contribution can be treated as the first step of the perturbative
expansion in the dimensionless parameter
$$
\kappa = \frac{l_{P}^{(N-1)/2}}{V^{(N-1)/2N}}.
$$
Indeed, due to definition of creation and annihilation operators
(\ref{g-via-a}), (\ref{pi-via-a}) in which momentum $\pi^{ij}$ is
proportional to $1/l_{P}^{(N-1)/2}$ while $g_{ij}$ is proportional
to $l_{P}^{(N-1)/2}$ the dependence on $l_{P}$ disappears from super-
Hamiltonian in quadratic approximation. $k$-th order of expansion in
the metric (super-Hamiltonian contains only quadratic terms in
momenta) is proportional to $\kappa^{k-2}$. Thus, the terms of $k$-th
order in creation and annihilation operators are proportional to the
same power of the expansion parameter $\kappa$. So, this expansion
in $a,a^{+},\bar{a},\bar{a}^{+}$ can be thought about as plausible for
the stationary Universe whose size is greater than Planck length.

From Eqs. (\ref{superhamiltonian-n1}),
(\ref{supermomentum}),
(\ref{L-define}) it follows 
that the contribution of graviton modes into the central extension of
the commutator $[L(\vec{n}),L(-\vec{n})]$ looks as follows:
\begin{eqnarray}                                                   
&&[L(\vec{n}),L(-\vec{n})]_{c.e.\ 
gravitons}=\frac{n_{i}}{|\vec{n}|}
\times\left(\frac{1}{4}G_{ab,a'b'}G_{cd,c'd'}
G^{ab,cd}E_{i}^{a'b',c'd',e'}n_{e'}\omega_{\vec{n}}\right.\nonumber  
\\ 
&&+\frac{n_{e}n_{f}n_{e'}}{2\omega_{\vec{n}}}
G_{ab,c'd'}G_{cd,a'b'}C^{ab,cd,ef}F_{i}^{a'b',c'd',e'}
\nonumber \\
&&+\sum_{0<\vec{k}<\vec{n}}\left\{\frac{1}{4}G_{ab,a'b'}G_{cd,c'd'}
G^{ab,cd}E_{i}^{a'b',c'd',e'}k_{e'}\omega_{\vec{k}}\right.\nonumber \\
&&+\frac{1}{4}G_{ab,a'b'}G_{cd,c'd'}
G^{ab,cd}F_{i}^{a'b',c'd',e'}k_{e'}\omega_{\vec{n}-\vec{k}}\nonumber 
\\ 
&&+\frac{1}{2}G_{ab,a'b'}G_{cd,c'd'}
C^{ab,cd,ef}E_{i}^{a'b',c'd',e'}\frac{k_{e}k_{f}k_{e'}}
{\omega_{\vec{n}-\vec{k}}}\nonumber \\
&&+\frac{1}{2}G_{ab,c'd'}G_{cd,a'b'}C^{ab,cd,ef}E_{i}^{a'b',c'd',e'}
\frac{k_{e}k_{f}(n-k)_{e'}}
{\omega_{\vec{k}}}\nonumber \\
&&+\frac{1}{2}G_{ab,a'b'}G_{cd,c'd'}
C^{ab,cd,ef}F_{i}^{a'b',c'd',e'}\frac{(n-k)_{e}(n-k)_{f}k_{e'}}
{\omega_{\vec{k}}}\nonumber \\
&&+\frac{1}{2}G_{ab,c'd'}G_{cd,a'b'}C^{ab,cd,ef}F_{i}^{a'b',c'd',e'}
\frac{k_{e}k_{f}k_{e'}}
{\omega_{\vec{k}}}\nonumber \\
&&+\frac{1}{2}G_{ab,a'b'}G_{cd,c'd'}
D^{ab,e,cd,f}E_{i}^{a'b',c'd',e'}\frac{k_{e}(n-k)_{f}(n-k)_{e'}}
{\omega_{\vec{k}}}\nonumber \\
&&+\frac{1}{2}G_{ab,c'd'}G_{cd,a'b'}D^{ab,e,cd,f}E_{i}^{a'b',c'd',e'}
\frac{(n-k)_{e}k_{f}(n-k)_{e'}}
{\omega_{\vec{k}}}\nonumber \\
&&+\frac{1}{2}G_{ab,a'b'}G_{cd,c'd'}
D^{ab,e,cd,f}F_{i}^{a'b',c'd',e'}\frac{k_{e}(n-k)_{f}k_{e'}}
{\omega_{\vec{k}}}\nonumber \\
&&\left.+\frac{1}{2}G_{ab,c'd'}G_{cd,a'b'}D^{ab,e,cd,f}F_{i}^{a'b',c'd',e'}
\frac{(n-k)_{e}k_{f}k_{e'}}
{\omega_{\vec{k}}}\right\}\nonumber \\
&&+\sum_{\vec{k}>0}\left\{\frac{1}{4}G_{ab,a'b'}G_{cd,c'd'}
G^{ab,cd}E_{i}^{a'b',c'd',e'}((n+k)_{e'}\omega_{\vec{n+k}}
-k_{e'}\omega_{\vec{k}})\right.\nonumber \\
&&+\frac{1}{4}G_{ab,a'b'}G_{cd,c'd'}
G^{ab,cd}F_{i}^{a'b',c'd',e'}(\omega_{\vec{k}} (n+k)_{e'}-
\omega_{\vec{n}+\vec{k}} k_{e'})\nonumber \\
&&+\frac{1}{2}G_{ab,a'b'}G_{cd,c'd'} C^{ab,cd,ef} E_{i}^{a'b',c'd',e'}
\left(\frac{(n+k)_{e}(n+k)_{f}(n+k)_{e'}}{\omega_{\vec{k}}}
-\frac{k_{e}k_{f}k_{e'}}{\omega_{\vec{n}+\vec{k}}}\right)\nonumber \\
&&+\frac{1}{2}G_{ab,c'd'}G_{cd,a'b'} C^{ab,cd,ef} E_{i}^{a'b',c'd',e'}
\left(\frac{k_{e}k_{f}(n+k)_{e'}}{\omega_{\vec{k}}}
-\frac{(n+k)_{e}(n+k)_{f}k_{e'}}
{\omega_{\vec{n}+\vec{k}}}\right)\nonumber \\
&&+\frac{1}{2}G_{ab,a'b'}G_{cd,c'd'} C^{ab,cd,ef} F_{i}^{a'b',c'd',e'}
\left(\frac{k_{e}k_{f}(n+k)_{e'}}{\omega_{\vec{n}+\vec{k}}}
-\frac{(n+k)_{e}(n+k)_{f}k_{e'}}{\omega_{\vec{k}}}\right)\nonumber \\
&&+\frac{1}{2}G_{ab,c'd'}G_{cd,a'b'} C^{ab,cd,ef} F_{i}^{a'b',c'd',e'}
\left(\frac{(n+k)_{e}(n+k)_{f}(n+k)_{e'}}{\omega_{\vec{n}+\vec{k}}}
-\frac{k_{e}k_{f}k_{e'}}{\omega_{\vec{k}}}\right)\nonumber \\
&&+\frac{1}{2}G_{ab,a'b'}G_{cd,c'd'} D^{ab,e,cd,f} 
E_{i}^{a'b',c'd',e'} 
\left(\frac{(n+k)_{e}k_{f}k_{e'}}{\omega_{\vec{n}+\vec{k}}}
-\frac{k_{e}(n+k)_{f}(n+k)_{e'}}{\omega_{\vec{k}}}\right)\nonumber \\
&&+\frac{1}{2}G_{ab,c'd'}G_{cd,a'b'} D^{ab,e,cd,f} 
E_{i}^{a'b',c'd',e'} 
\left(\frac{k_{e}(n+k)_{f}k_{e'}}{\omega_{\vec{n}+\vec{k}}}
-\frac{(n+k)_{e}k_{f}(n+k)_{e'}}{\omega_{\vec{k}}}\right)\nonumber \\
&&+\frac{1}{2}G_{ab,a'b'}G_{cd,c'd'} D^{ab,e,cd,f} 
F_{i}^{a'b',c'd',e'} 
\left(\frac{k_{e}(n+k)_{f}k_{e'}}{\omega_{\vec{k}}}
-\frac{(n+k)_{e}k_{f}(n+k)_{e'}}{\omega_{\vec{n}+\vec{k}}}\right)
\nonumber \\
&&+\frac{1}{2}G_{ab,c'd'}G_{cd,a'b'} D^{ab,e,cd,f} 
F_{i}^{a'b',c'd',e'}\nonumber \\ 
&&\left.\left.\times\left(\frac{(n+k)_{e}k_{f}k_{e'}}{\omega_{\vec{k}}}
-\frac{k_{e}(n+k)_{f}(n+k)_{e'}}{\omega_{\vec{n}+\vec{k}}}\right)
\right\}\right).
\label{central-gravitons}
\end{eqnarray}

It is easy to get the contractions of structure tensors $G, C, D, E$
and $F$ defined by formulae (\ref{supermetric}),(\ref{supermetric1}),
(\ref{E-define})--(\ref{D-define}). These contractions look as
follows:
\begin{equation}
G_{ab,a'b'}G_{cd,c'd'}G^{ab,cd}F_{i}^{a'b',c'd',e'} = 
\frac{(N-2)(N+1)}{2} \delta_{i}^{e'},
\label{G-F-contr}
\end{equation}
\begin{eqnarray}
&&G_{ab,a'b'}G_{cd,c'd'}
C^{ab,cd,ef}E_{i}^{a'b',c'd',e'} = 
G_{ab,c'd'}G_{cd,a'b'}C^{ab,cd,ef}E_{i}^{a'b',c'd',e'}\nonumber \\
&&=(N-2)(N+1)(\delta_{i}^{e'}\delta^{ef} -
\delta_{i}^{(e}\delta^{f)e'}),
\label{C-E-contr}
\end{eqnarray}
\begin{eqnarray}
&&G_{ab,a'b'}G_{cd,c'd'}
C^{ab,cd,ef}F_{i}^{a'b',c'd',e'} =
G_{ab,c'd'}G_{cd,a'b'}C^{ab,cd,ef}F_{i}^{a'b',c'd',e'}\nonumber \\
&&=-\frac{(N-2)(N+1)}{2(N-1)}((N-3)\delta_{i}^{e'}\delta^{ef}
+2\delta_{i}^{(e}\delta^{f)e'}),
\label{C-F-contr}
\end{eqnarray}
\begin{eqnarray}
&&G_{ab,a'b'}G_{cd,c'd'}D^{ab,e,cd,f}E_{i}^{a'b',c'd',e'}=
\frac{3N^{2} - 3N - 8}{4(N-1)}\delta_{i}^{e'}\delta^{ef}\nonumber \\
&&-\frac{N^{2}-N-4}{4(N-1)}\delta_{i}^{f}\delta^{ee'} -
\frac{N^{2}-3}{2(N-1)}\delta_{i}^{e}\delta^{fe'},
\label{D-E-contr}
\end{eqnarray}
\begin{eqnarray}
&&G_{ab,c'd'}G_{cd,a'b'}D^{ab,e,cd,f}E_{i}^{a'b',c'd',e'}=
\frac{3N^{2} - 3N - 8}{4(N-1)}\delta_{i}^{e'}\delta^{ef}\nonumber \\
&&-\frac{N^{2}-N-4}{4(N-1)}\delta_{i}^{e}\delta^{e'f} -
\frac{N^{2}-3}{2(N-1)}\delta_{i}^{f}\delta^{ee'},
\label{D-E-contr1}
\end{eqnarray}
\begin{eqnarray}
&&G_{ab,a'b'}G_{cd,c'd'}D^{ab,e,cd,f}F_{i}^{a'b',c'd',e'}=
\frac{-3N^{3}+6N^{2}+N-20}{8(N-1)}\delta_{i}^{e'}\delta{ef}
\nonumber \\
&&-\frac{N^{2}-9}{8(N-1)}\delta_{i}^{e}\delta^{e'f}
+\frac{N^{2}+2N+5}{8(N-1)}\delta_{i}^{f}\delta^{ee'},
\label{D-F-contr}
\end{eqnarray}
\begin{eqnarray}
&&G_{ab,c'd'}G_{cd,a'b'}D^{ab,e,cd,f}F_{i}^{a'b',c'd',e'}=
\frac{-3N^{3}+6N^{2}+N-20}{8(N-1)}\delta_{i}^{e'}\delta{ef}
\nonumber \\
&&-\frac{N^{2}-9}{8(N-1)}\delta_{i}^{f}\delta^{ee'}
+\frac{N^{2}+2N+5}{8(N-1)}\delta_{i}^{e}\delta^{fe'}.
\label{D-F-contr1}
\end{eqnarray}

In the ``last-element limit" (see sec.3), the expression
(\ref{central-gravitons}) is reduced to the much more simple one
\begin{eqnarray}
&&[L(\vec{n}),L(-\vec{n})]_{c.e.\ 
gravitons}=\nonumber \\
&&(GGGE)_{\rm l.e.l.}\frac{1}{4}\left(n^{2} + 
\sum_{k=1}^{n-1} k^{2} - \sum_{k=1}^{\infty} 
k^{2}+\sum_{k=n+1}^{\infty} k^{2}\right)\nonumber \\
&&+(GGGF)_{\rm l.e.l.} \left(\frac{1}{4}\sum_{k=1}^{n-1} k 
(n-k)\right) \nonumber \\
&&+(GGCE)_{\rm l.e.l.}\left(\frac{1}{2}\sum_{k=1}^{n-1}\left(
\frac{k^{3}}{n-k} + k(n-k)\right)\right.\nonumber \\
&&\left.+\frac{1}{2}\sum_{k=1}^{\infty}\left(-\frac{k^{3}}{n+k}
+\frac{(n+k)^{3}}{k}\right)\right)\nonumber \\
&&+(GGCF)_{\rm l.e.l.}\left(\frac{1}{2}n^{2} +
\sum_{k=1}^{n-1} k^{2}\right)\nonumber \\
&&+(GGDE)_{\rm l.e.l.}\left(
\sum_{k=1}^{n-1} k^{2} + \sum_{k=1}^{\infty} 
k^{2}-\sum_{k=n+1}^{\infty} k^{2}\right)\nonumber \\
&&+(GGDF)_{\rm l.e.l.}\left(\sum_{k=1}^{n-1} 
k(n-k)\right).
\label{central-gravitons1}
\end{eqnarray}
Here the notations $(GGGE)_{\rm l.e.l.}$ etc. denote the 
corresponding contractions taken in the ``last-element'' limit i.e. 
in the case when $e=f=e'=N$.
It is easy
to see that the coefficient at $(GGGE)_{\rm l.e.l.}$ in Eq. 
(\ref{central-gravitons1}) vanishes. The structure at $(GGCE)_{\rm 
l.e.l.}$ is very inconvenient for calculation, but fortunately 
$$
(GGGE)_{\rm l.e.l.} = 0.
$$
Other sums can be
calculated due to simple formulae
\begin{equation}
\sum_{k=1}^{n-1} 
k(n-k) = \frac{n(n^{2}-1)}{6},
\label{sum1}
\end{equation} 
\begin{equation}
\sum_{k=1}^{n-1} k^{2} = \frac{n(n-1)(2n-1)}{6},
\label{sum2}
\end{equation}
\begin{equation}
\sum_{k=1}^{n} k^{2} = \frac{n(n+1)(2n+1)}{6}.
\label{sum3}
\end{equation}
Using Eqs. (\ref{sum1})--(\ref{sum3}) one reduces Eq.
(\ref{central-gravitons1}) to
\begin{eqnarray}
&&[L(\vec{n}),L(-\vec{n})]_{c.e.\
gravitons}=\nonumber \\
&&n^{3}\left(\frac{1}{24}(GGGF)_{\rm l.e.l.} + \frac{1}{3}(GGCF)_{\rm 
l.e.l.} + \frac{2}{3}(GGDE)_{\rm l.e.l.} + \frac{1}{6}(GGDF)_{\rm 
l.e.l.}\right)\nonumber \\
&&+n\left(-\frac{1}{24}(GGGF)_{\rm l.e.l.} + \frac{1}{6}(GGCF)_{\rm 
l.e.l.} + \frac{1}{3}(GGDE)_{\rm l.e.l.} - \frac{1}{6}(GGDF)_{\rm 
l.e.l.}\right).
\label{central-gravitons2}
\end{eqnarray}
We shall be interested mainly in the coefficient at $n^{3}$ (because as it
is well-known the coefficients at $n$ can be adjusted to the 
necessary  values by inclusion of intercept into $L_{0}$), however
we shall write down the full expression for the graviton contribution
into the quantum correction to $[L(\vec{n}),L(-\vec{n})]$.
Now we shall calculate
(\ref{central-gravitons2} using the following formulae for the
contractions in the ``last-element'' limit:
\begin{equation}
(GGGF)_{\rm l.e.l.} = \frac{(N-2)(N+1)}{2},
\label{contraction1}
\end{equation}
\begin{equation}
(GGCF)_{\rm l.e.l.} = -\frac{(N-2)(N+1)}{2},
\label{contraction2}
\end{equation}
\begin{equation}
(GGDEF)_{\rm l.e.l.} = -\frac{1}{2},
\label{contraction3}
\end{equation}
\begin{equation}
(GGDF)_{\rm l.e.l.} = -\frac{3(N-2)(N+1)}{8}.
\label{contraction4}
\end{equation}
Substituting Eqs. (\ref{contraction1})--(\ref{contraction4}) 
into Eq. (\ref{central-gravitons2}) we have
\begin{eqnarray}
&&[L(\vec{n}),L(-\vec{n})]_{c.e.\
gravitons} =
-\frac{1}{12} \left(4 + \frac{5}{2} (N-2)(N+1)\right) n^{3}\nonumber\\
&&-\frac{1}{12} \left(2 +\frac{1}{2} (N-2)(N+1)\right) n.
\label{central-gravitons3}
\end{eqnarray}
Notice, that the appearance of the expression
$(N-2)(N+1)/2$ in the right-hand side of Eq. (\ref{central-gravitons3})
does not look surprising because this expression is nothing but the
number of gravitons. What is less trivial that the coefficient
at the number of gravitons differs from the one at the number of matter
fields and even has another sign. Moreover, there is a constant
contribution in the right-hand side of Eq. (\ref{central-gravitons3}),
which is independent from the dimensionality of space.

Concluding this section we write the contributions of some other
matter fields into the central extension. The contribution of massless
vector fields is quite analogous of that of scalar field. The number of
local degrees of freedom of one vector field is $(N-1)$. Thus, if the model
includes $d_{V}$ vector fields their contribution is
\begin{equation}
[L(\vec{n}),L(-\vec{n})]_{c.e.\ vectors} =
\frac{d_{V} (N-1)}{12} n (n^{2} - 1).
\label{central-vectors}
\end{equation}

The Weyl or Majorana massless spinor field has $2^{[(N+1)/2 -1]}$
local degrees of freedom
\footnote{It is interesting to recall that in the space--times of the
exceptional dimensionality $2+8k$, with integer $k$, one can satisfy
simultaneously
Weyl and Majorana conditions and the number of degrees of freedom of spinor
is $2^{(\frac{N+1}{2} - 2})$ for this case. For example, in the simplest
$1+1$--dimensional case (i.e. for $N=1$) the contribution of the spinor should
be one--half of scalar contribution (see e.g. \cite{Goddard-Olive}).}
and the contribution of such fields
into $[L(\vec{n}),L(-\vec{n})]_{c.e.}$ looks as
\begin{equation}
[L(\vec{n}),L(-\vec{n})]_{c.e. spinors} =
\frac{d_{F}}{12} 2^{[(N+1)/2 -1]} n (n^{2} - 1).
\label{central-spinor}
\end{equation}

Let us collect the contributions of all the matter fields and gravity
into the central extension $[L(\vec{n}),L(\vec{-n})]$:
\begin{eqnarray}
&&[L(\vec{n}),L(-\vec{n})]_{c.e.} = \nonumber\\
&&\frac{d}{12} n (n^{2} - 1)
+ \frac{d_{V} (N-1)}{12} n (n^{2} - 1)
+ \frac{d_{F}}{12} 2^{[(N+1)/2 -1]} n (n^{2} - 1)\nonumber\\
&&-\frac{1}{12} \left(4 +\frac{5}{2} (N-2)(N+1)\right) n^{3}
-\frac{1}{12} \left(2 +\frac{1}{2} (N-2)(N+1)\right) n
\label{central-matter}
\end{eqnarray}

One can show by direct calculation that the commutators between area-
preserving diffeomorphisms and between them and Virasoro-like generators
do not contain quantum corrections. The commutators between $L$ and
$\bar{L}$ Virasoro-like generators do not include
first quantum corrections
as well. It is easy to see that the quantum correction to the
commutator $[\bar{L}(-\vec{n}),\bar{L}(\vec{n})]$ coincides with one
for $[L(\vec{n}),L(-\vec{n})]$ given by the right-hand side of Eq.
(\ref{central-matter}).

Thus we have seen that the algebra of constraints is disclosed already by
constant contributions of the first quantum corrections. It is worth mentioning
that the fact of a disclosure does not mean by itself the quantum inconsistency
of the theory. A proper test for consistency is to examine the nilpotency
condition for the quantum BRST charge related to the disclosed constraint
algebra. This examination is performed in the next section.

\section{BRST operator and critical relation \newline
between parameters of the theory}

In this section we study the operatorial canonical BFV quantization
\cite{BF-operator,BF-86,Bat-spur} of the constrained
dynamics developed in the
previous section for the toroidal Bianchi-I cosmology. The general
prescription
of the BFV--method implies making a definite choice of the operator
symbol both
for the original phase space variables and ghosts. The main insight at the
choice
of the ghost ordering is from the due regard to
the structure of the related constraints.
The Virasoro--like subset of the constraints has the polarization,
splitting them
into the positive-- and negative--frequency ones,
being conjugated to each other.
Thereby, it is natural to subject the related ghost variables to the Wick
ordering rule. Meanwhile the ghosts attached to the area--preserving
diffeomorphisms are quantized by the Weyl rule. There are several reasons for
such a choice of ordering for these ghosts. On the one hand, this ordering,
as it will be shown later, is compatible with the mentioned in previous
section fact, that there are no quantum corrections to the commutators
of the area--preserving diffeomorphisms among themselves and with
the Virasoro--like constraints. On the other hand, this choice follows to
some extent to a certain analogy with $p$--brane theory \cite{area-preserving}
where the area preserving constraints describe some residual symmetry after
fixing of the light--cone gauge. Now we are in a position to write down
the quantum BRST charge for the model. It looks like
\begin{eqnarray}
&&\Omega = \sum_{\vec{n}>0} (c^{+(\vec{n})}L(\vec{n})
+ c^{(\vec{n})}L(-\vec{n}) + \bar{c}^{(\vec{n})}\bar{L}(\vec{n})
+\bar{c}^{+(\vec{n})}\bar{L}(-\vec{n}))\nonumber\\
&&+c^{(0)}L(0) + \bar{c}^{(0)}\bar{L}(0)
+\sum_{\vec{n},\vec{v}}\tilde{c}^{(\vec{n},\vec{v})}
\tilde{H}_{\vec{v}}(\vec{n})\nonumber\\
&&+\frac{1}{2}\sum_{\vec{n}>0,\vec{m}>0}c^{+(\vec{m})}
c^{+(\vec{n})}P_{(\vec{n}+\vec{m})} U_{L(\vec{n}) L(\vec{m})}
^{L(\vec{n}+\vec{m})}
+\frac{1}{2}\sum_{\vec{n}>0,\vec{m}>0}c^{+(\vec{m})}
c^{+(\vec{n})}\bar{P}^{+}_{(\vec{n}+\vec{m})} U_{L(\vec{n}) L(\vec{m})}
^{\bar{L}(\vec{n}+\vec{m})}\nonumber\\
&&+\frac{i}{2}\sum_{\vec{n}>0,\vec{m}>0,\vec{v}}c^{+(\vec{m})}
c^{+(\vec{n})}\tilde{P}_{(\vec{n}+\vec{m},\vec{v})} U_{L(\vec{n}) L(\vec{m})}
^{\tilde{H}(\vec{n}+\vec{m},\vec{v})}
+\frac{1}{2}\sum_{\vec{n}<0,\vec{m}<0}\bar{c}^{+(\vec{m})}
\bar{c}^{+(\vec{n})}\bar{P}_{(\vec{n}+\vec{m})} U_{\bar{L}(\vec{n})
\bar{L}(\vec{m})}
^{\bar{L}(\vec{n}+\vec{m})}\nonumber\\
&&+\frac{1}{2}\sum_{\vec{n}<0,\vec{m}<0}\bar{c}^{+(\vec{m})}
\bar{c}^{+(\vec{n})}P^{+}_{(\vec{n}+\vec{m})} U_{\bar{L}(\vec{n})
\bar{L}(\vec{m})}
^{L(\vec{n}+\vec{m})}
+\frac{i}{2}\sum_{\vec{n}<0,\vec{m}<0,\vec{v}}\bar{c}^{+(\vec{m})}
\bar{c}^{+(\vec{n})}\tilde{P}_{(\vec{n}+\vec{m},\vec{v})}
U_{\bar{L}(\vec{n}) \bar{L}(\vec{m})}^{\tilde{H}(\vec{n}+\vec{m},\vec{v})}
\nonumber\\
&&+\sum_{\vec{n}>0,\vec{n}>\vec{m}>0}c^{+(\vec{n})}
P_{(\vec{n}-\vec{m})}c^{(\vec{m})}
U_{L(\vec{n})L(-\vec{m})}^{L(\vec{n}-\vec{m})}
+\sum_{\vec{n}>0,\vec{n}>\vec{m}>0}c^{+(\vec{n})}
\bar{P}^{+}_{(\vec{n}-\vec{m})}c^{(\vec{m})}
U_{L(\vec{n})L(-\vec{m})}^{\bar{L}(\vec{n}-\vec{m})}\nonumber\\
&&+i\sum_{\vec{n}>0,\vec{n}>\vec{m}>0,\vec{v}}c^{+(\vec{n})}
c^{(\vec{m})}\tilde{P}^{+}_{(\vec{n}-\vec{m},\vec{v})}
U_{L(\vec{n})L(-\vec{m})}^{\tilde{H}(\vec{n}-\vec{m},\vec{v})}
+\sum_{\vec{n}<0,\vec{n}<\vec{m}<0}\bar{c}^{+(\vec{n})}
\bar{P}_{(\vec{n}-\vec{m})}\bar{c}^{(\vec{m})}
U_{\bar{L}(\vec{n})\bar{L}(-\vec{m})}^{\bar{L}(\vec{n}-\vec{m})}\nonumber\\
&&+\sum_{\vec{n}<0,\vec{n}<\vec{m}<0}\bar{c}^{+(\vec{n})}
P^{+}_{(\vec{n}-\vec{m})}\bar{c}^{(\vec{m})}
U_{\bar{L}(\vec{n})\bar{L}(-\vec{m})}^{L(\vec{n}-\vec{m})}
+i\sum_{\vec{n}<0,\vec{n}<\vec{m}<0,\vec{v}}\bar{c}^{+(\vec{n})}
\bar{c}^{(\vec{m})}\tilde{P}^{+}_{(\vec{n}-\vec{m},\vec{v})}
U_{\bar{L}(\vec{n})\bar{L}(-\vec{m})}
^{\tilde{H}(\vec{n}-\vec{m},\vec{v})}\nonumber\\
&&+i\sum_{\vec{n}>0}c^{+(\vec{n})}P_{(0)}c^{(\vec{n})}
U_{L(\vec{n}) L(-\vec{n})}^{L(0)}
+\sum_{\vec{n}>0}c^{+(\vec{n})}c^{(0)}P_{(\vec{n})}
U_{L(\vec{n})L(0)}^{L(\vec{n})}\nonumber\\
&&+i\sum_{\vec{n}<0}\bar{c}^{+(\vec{n})}\bar{P}_{(0)}\bar{c}^{(\vec{n})}
U_{\bar{L}(\vec{n}) \bar{L}(-\vec{n})}^{\bar{L}(0)}
+\sum_{\vec{n}<0}\bar{c}^{+(\vec{n})}\bar{c}^{(0)}\bar{P}_{(\vec{n})}
U_{\bar{L}(\vec{n})\bar{L}(0)}^{\bar{L}(\vec{n})}\nonumber\\
&&+\sum_{\vec{n}>0,\vec{m}>0}c^{+(\vec{n})}P_{(\vec{n}+\vec{m})}
\bar{c}^{(\vec{m})} U_{L(\vec{n}) \bar{L}(\vec{m})}
^{L(\vec{n}+\vec{m})}
+\sum_{\vec{n}>0,\vec{m}>0}c^{+(\vec{n})}\bar{P}^{+}_{(\vec{n}+\vec{m})}
\bar{c}^{(\vec{m})} U_{L(\vec{n}) \bar{L}(\vec{m})}
^{\bar{L}(\vec{n}+\vec{m})}\nonumber\\
&&+i\sum_{\vec{n}>0,\vec{m}>0,\vec{v}}c^{+(\vec{n})}
\tilde{P}_{(\vec{n}+\vec{m},\vec{v})}
\bar{c}^{(\vec{m})} U_{L(\vec{n}) \bar{L}(\vec{m})}
^{\tilde{H}(\vec{n}+\vec{m},\vec{v})}
+\sum_{\vec{n}>0,\vec{n}>\vec{m}>0}\bar{c}^{+(\vec{m})}
c^{+(\vec{n})}P_{(\vec{n}-\vec{m})}
U_{L(\vec{n}) \bar{L}(-\vec{m})}
^{L(\vec{n}-\vec{m})}\nonumber\\
&&+\sum_{\vec{n}>0,\vec{n}>\vec{m}>0}\bar{c}^{+(\vec{m})}
c^{+(\vec{n})}\bar{P}^{+}_{(\vec{n}-\vec{m})}
U_{L(\vec{n}) \bar{L}(-\vec{m})}
^{\bar{L}(\vec{n}-\vec{m})}
+i\sum_{\vec{n}>0,\vec{n}>\vec{m}>0,\vec{v}}\bar{c}^{+(\vec{m})}
c^{+(\vec{n})}\tilde{P}_{(\vec{n}-\vec{m},\vec{v})}
U_{L(\vec{n}) \bar{L}(-\vec{m})}
^{\tilde{H}(\vec{n}-\vec{m},\vec{v})}\nonumber\\
&&+\sum_{\vec{m}>0,\vec{m}>\vec{n}>0}\bar{c}^{+(\vec{m})}
c^{+(\vec{n})}P^{+}_{(\vec{n}-\vec{m})}
U_{L(\vec{n}) \bar{L}(-\vec{m})}
^{L(\vec{n}-\vec{m})}
+\sum_{\vec{m}>0,\vec{m}>\vec{n}>0}\bar{c}^{+(\vec{m})}
c^{+(\vec{n})}\bar{P}_{(\vec{n}-\vec{m})}
U_{L(\vec{n}) \bar{L}(-\vec{m})}
^{\bar{L}(\vec{n}-\vec{m})}\nonumber\\
&&+i\sum_{\vec{m}>0,\vec{m}>\vec{n}>0,\vec{v}}\bar{c}^{+(\vec{m})}
c^{+(\vec{n})}\tilde{P}_{(\vec{n}-\vec{m},\vec{v})}
U_{L(\vec{n}) \bar{L}(-\vec{m})}
^{\tilde{H}(\vec{n}-\vec{m},\vec{v})}
+\sum_{\vec{n}>0,\vec{m}>-\vec{n}}
c^{+(\vec{n})}\tilde{c}^{(\vec{m},\vec{v})}P_{(\vec{n}+\vec{m})}
U_{L(\vec{n}) \tilde{H}(\vec{m},\vec{v})}^{L(\vec{n}+\vec{m})}\nonumber\\
&&+\sum_{\vec{n}>0,\vec{m}>-\vec{n}}
\bar{P}^{+}_{(\vec{n}+\vec{m})}
c^{+(\vec{n})}\tilde{c}^{(\vec{m},\vec{v})}
U_{L(\vec{n}) \tilde{H}(\vec{m},\vec{v})}^{\bar{L}
(\vec{n}+\vec{m})}
+i\sum_{\vec{n}>0,\vec{m}>-\vec{n}}
c^{+(\vec{n})}\tilde{c}^{(\vec{m},\vec{v})}
\tilde{P}_{(\vec{n}+\vec{m},\vec{w})}
U_{L(\vec{n}) \tilde{H}(\vec{m},\vec{v})}
^{\tilde{H}(\vec{n}+\vec{m},\vec{w})}\nonumber\\
&&+\sum_{\vec{n}<0,\vec{m}>\vec{n}}
\bar{c}^{+(\vec{n})}\tilde{c}^{(\vec{m},\vec{v})}\bar{P}_{(\vec{n}+\vec{m})}
U_{\bar{L}(\vec{n}) \tilde{H}(\vec{m},\vec{v})}^{\bar{L}
(\vec{n}+\vec{m})}
+\sum_{\vec{n}<0,\vec{m}>\vec{n}}
P^{+}_{(\vec{n}+\vec{m})}
\bar{c}^{+(\vec{n})}\tilde{c}^{(\vec{m},\vec{v})}
U_{\bar{L}(\vec{n}) \tilde{H}(\vec{m},\vec{v})}^{L
(\vec{n}+\vec{m})}\nonumber\\
&&+i\sum_{\vec{n}<0,\vec{m}>\vec{n}}
\bar{c}^{+(\vec{n})}\tilde{c}^{(\vec{m},\vec{v})}
\tilde{P}_{(\vec{n}+\vec{m},\vec{w})}
U_{\bar{L}(\vec{n}) \tilde{H}(\vec{m},\vec{v})}
^{\tilde{H}(\vec{n}+\vec{m},\vec{w})}
+i\sum_{\vec{n}>0}c^{+(\vec{n})}\tilde{c}^{(-\vec{n},\vec{v})}
\tilde{P}_{(0,\vec{v})} U_{L(\vec{n}) \tilde{H}(-\vec{n},\vec{v})}
^{\tilde{H}(0,\vec{v})}\nonumber\\
&&+i\sum_{\vec{n}<0}\bar{c}^{+(\vec{n})}\tilde{c}^{(-\vec{n},\vec{v})}
\tilde{P}_{(0,\vec{v})} U_{\bar{L}(\vec{n}) \tilde{H}(-\vec{n},\vec{v})}
^{\tilde{H}(0,\vec{v})}
+\sum_{\vec{n}>0,\vec{m}<-\vec{n}}
P^{+}_{(\vec{n}+\vec{m})}c^{+(\vec{n})}\tilde{c}^{(\vec{m},\vec{v})}
U_{L(\vec{n}) \tilde{H}(\vec{m},\vec{v})}^{L(\vec{n}+\vec{m})}\nonumber\\
&&+\sum_{\vec{n}>0,\vec{m}<-\vec{n}}
c^{+(\vec{n})}\tilde{c}^{(\vec{m},\vec{v})}
\bar{P}_{(\vec{n}+\vec{m})}
U_{L(\vec{n}) \tilde{H}(\vec{m},\vec{v})}
^{\bar{L}(\vec{n}+\vec{m})}
+i\sum_{\vec{n}>0,\vec{m}<-\vec{n}}
c^{+(\vec{n})}\tilde{c}^{(\vec{m},\vec{v})}
\tilde{P}_{(\vec{n}+\vec{m})}
U_{L(\vec{n}) \tilde{H}(\vec{m},\vec{v})}
^{\tilde{H}(\vec{n}+\vec{m},\vec{w})}\nonumber\\
&&+\sum_{\vec{n}<0,\vec{m}>-\vec{n}}
\bar{P}^{+}_{(\vec{n}+\vec{m})}
\bar{c}^{+(\vec{n})}\tilde{c}^{(\vec{m},\vec{v})}
U_{\bar{L}(\vec{n}) \tilde{H}(\vec{m},\vec{v})}^{
\bar{L}(\vec{n}+\vec{m})}
+\sum_{\vec{n}<0,\vec{m}>-\vec{n}}
\bar{c}^{+(\vec{n})}\tilde{c}^{(\vec{m},\vec{v})}
P_{(\vec{n}+\vec{m})}
U_{\bar{L}(\vec{n}) \tilde{H}(\vec{m},\vec{v})}
^{L(\vec{n}+\vec{m})}\nonumber\\
&&+i\sum_{\vec{n}<0,\vec{m}>-\vec{n}}
\bar{c}^{+(\vec{n})}\tilde{c}^{(\vec{m},\vec{v})}
\tilde{P}_{(\vec{n}+\vec{m})}
U_{\bar{L}(\vec{n}) \tilde{H}(\vec{m},\vec{v})}
^{\tilde{H}(\vec{n}+\vec{m},\vec{w})}\nonumber\\
&&+h.c.\nonumber\\
&&+\frac{1}{6}\sum_{\vec{n},\vec{m},\vec{v},\vec{w}}
(\tilde{c}^{(\vec{m},\vec{w})}\tilde{c}^{(\vec{m},\vec{w})}
\tilde{P}_{(\vec{n}+\vec{m},\vec{u})}
+\tilde{P}_{(\vec{n}+\vec{m},\vec{u})}
\tilde{c}^{(\vec{m},\vec{w})}\tilde{c}^{(\vec{m},\vec{w})}\nonumber\\
&&+\tilde{c}^{(\vec{m},\vec{w})}
\tilde{P}_{(\vec{n}+\vec{m},\vec{u})}
\tilde{c}^{(\vec{m},\vec{w})})U_{\tilde{H}(\vec{n},\vec{v})
\tilde{H}(\vec{m},\vec{w})}^{\tilde{H}(\vec{n}+\vec{m},\vec{u})}.
\label{BRST}
\end{eqnarray}
In Eq. (\ref{BRST}) we should make summation over the set of
$(N-1)$ unit
vectors $\vec{v}$ enumerating area-preserving diffeomorphisms
constraints $\tilde{H}_{\vec{v}}^{(\vec{n})}$ which are orthogonal
to vector $\vec{n}$.

The ghosts $c^{+(\vec{n})}, \bar{c}^{+(\vec{n})}, P^{+}_{(\vec{n})}$
and $\bar{P}^{+}_{(\vec{n})}$ are creation operators,
$c^{(\vec{n})}, \bar{c}^{(\vec{n})}, P_{(\vec{n})}$
and $\bar{P}_{(\vec{n})}$ are annihilation operators.
All these operators are Wick-ordered,
while
$\tilde{c}^{(\vec{n},\vec{v})}, \tilde{P}_{(\vec{n},\vec{v})},
c^{(0)}, P_{(0)}, \bar{c}^{(0)}$ and $\bar{P}_{(0)}$ are ordered
according to Weyl rule. The latter are subjected the following
nonvanishing anticommutation relations:
\begin{eqnarray}
&&[c^{+(\vec{n})}, P_{(\vec{m})}] = \delta^{\vec{(n)}}_{(\vec{m})},
\nonumber\\
&&[c^{(\vec{n})}, P^{+}_{(\vec{m})}] = \delta^{\vec{(n)}}_{(\vec{m})},
\nonumber\\
&&[\bar{c}^{+(\vec{n})}, \bar{P}_{(\vec{m})}] = \delta^{\vec{(n)}}_{(\vec{m})},
\nonumber\\
&&[\bar{c}^{(\vec{n})}, \bar{P}^{+}_{(\vec{m})}] =
\delta^{\vec{(n)}}_{(\vec{m})},
\\
&&[\tilde{c}^{(\vec{n},\vec{v})}\tilde{P}_{(\vec{m},\vec{w})}]
= i\delta^{\vec{(n)}}_{(\vec{m})} \delta^{(\vec{v}-\vec{w})}_{0},\nonumber\\
&&[c^{0}, P_{0}] = i,
\nonumber\\
&&[\bar{c}^{0}, \bar{P}_{0}] = i. \nonumber
\label{anticommutators}
\end{eqnarray}

Making use of Wick theorem, one may establish that due to the terms with two
contractions of ghost operators,
the requirement of vanishing of the squared BRST
operator ${\Omega}^2=0$ should give rise to the following
nonvanishing corrections to the quantum
involution relations of the constraint operators:
\begin{eqnarray}
&&[L(\vec{n}),L(-\vec{n})]_{c.e.\;ghosts}=\frac{1}{2}
U_{L(\vec{n})\;L(0)}^{L(\vec{n})}
U_{L(\vec{n})\;L(-\vec{n})}^{L(0)}
+\frac{1}{2}
U_{L(\vec{n})\;L(-\vec{n})}^{L(0)}
U_{L(0)\;L(-\vec{n})}^{L(-\vec{n})}
\nonumber \\
&&+\sum_{0<\vec{k}<\vec{n}} 
U_{L(\vec{n})\;L(-\vec{k})}^{L(\vec{n}-\vec{k})}
U_{L(\vec{n}-\vec{k})\;L(-\vec{n})}^{L(-\vec{k})}\nonumber \\
&&+\sum_{\vec{k}>0}
(U_{L(\vec{n})\;\bar{L}(\vec{k})}^{L(\vec{n}+\vec{k})}
U_{L(\vec{n}+\vec{k})\;L(-\vec{n})}^{\bar{L}(\vec{k})}
+U_{L(\vec{n})\;L(-\vec{n}-\vec{k})}^{\bar{L}(-\vec{k})}
U_{\bar{L}(-\vec{k})\;L(-\vec{n})}^{L(-\vec{n}-\vec{k})});
\label{central-ghosts0}
\end{eqnarray}
\begin{eqnarray}
&&[\bar{L}(-\vec{n}),\bar{L}(\vec{n})]_{c.e.\;ghosts}=\frac{1}{2}
U_{\bar{L}(-\vec{n})\;\bar{L}(0)}^{\bar{L}(-\vec{n})}
U_{\bar{L}(-\vec{n})\;\bar{L}(\vec{n})}^{\bar{L}(0)}
+\frac{1}{2}
U_{\bar{L}(-\vec{n})\;\bar{L}(\vec{n})}^{\bar{L}(0)}
U_{\bar{L}(0)\;\bar{L}(\vec{n})}^{\bar{L}(\vec{n})}
\nonumber \\
&&+\sum_{0<\vec{k}>-\vec{n}}
U_{\bar{L}(-\vec{n})\bar{L}(-\vec{k})}^{\bar{L}(-\vec{n}-\vec{k})}
U_{\bar{L}(-\vec{n}-\vec{k})\;\bar{L}(\vec{n})}^{\bar{L}(-\vec{k})}
\nonumber \\
&&+\sum_{\vec{k}<0}
(U_{\bar{L}(-\vec{n})\;L(\vec{k})}^{\bar{L}(-\vec{n}+\vec{k})}
U_{\bar{L}(-\vec{n}+\vec{k})\;\bar{L}(\vec{n})}^{L(\vec{k})}
+U_{\bar{L}(-\vec{n})\;\bar{L}(\vec{n}-\vec{k})}^{L(\vec{k})}
U_{\bar{L}(-\vec{k})\;L(-\vec{n})}^{L(-\vec{n}-\vec{k})});
\label{central-ghosts1}
\end{eqnarray}

Straightforward but tedious consideration shows that the nonvanishing
quantum corrections arise only in commutators $[L(\vec{n}),L(-\vec{n})]$
and $[\bar{L}(-\vec{n}),\bar{L}(\vec{n})]$. Thus,
the rest of the commutators do not get $c$-number quantum corrections.
Moreover, all the contributions to Eqs. (\ref{central-ghosts0}),
(\ref{central-ghosts1}) of structure constants involving area-
preserving diffeomorphisms are mutually canceled.

Left--hand sides of the relations (\ref{central-ghosts0}) and
(\ref{central-ghosts1}) represent the quantum
corrections to the corresponding commutators calculated in the preceding
section, while the right-hand sides consist of the quantum ghost contributions
required by the nilpotency condition.
\footnote{The general fact of appearance the corrections which emerge from
the quantum contribution of ghost commutators has been first studied
in the paper \cite{Bat-spur}.}
Left--hand sides are determined by a
specific realization of the constraint algebra and, in particular, depend upon
the specific spectrum of the fields involved into the original Lagrangian,
whereas the right-hand sides do not depend on the specific realization
of the constraints, being completely fixed only by the structure constants
of the algebra. So, these equalities, following the nilpotency of the quantum
BRST charge, may constitute in principle the nontrivial restrictions on the
theory parameters.

It was shown in the preceding section non-trivial corrections
to quantum involution relations (see Eq. (\ref{central-matter})
depending on matter content of the theory appear only in commutators
$[L(\vec{n}),L(-\vec{n})]$ (and, symmetrically, in
$[\bar{L}(-\vec{n}),\bar{L}(\vec{n})]$). Here, we have an analogous
structure of quantum corrections to involution relations being originated
from quantum commutators between the terms in $\Omega$ which are of
third order in ghost variables. In this approximation, it is sufficient to
keep only the constant parts of the structure coefficients because the
multipole harmonics may contribute into the central extension only
if the terms with three or more contractions are accounted for. These
terms represent the quantum corrections of a higher order.

Now we should calculate the right-hand side of Eq.
(\ref{central-ghosts0}) explicitly (calculation of
(\ref{central-ghosts1}) does not give us an additional information).

Substituting into Eq. (\ref{central-ghosts0}) explicit expressions for
structure constant we have:
\begin{eqnarray}
&&[L(\vec{n}),L(-\vec{n})]_{c.e.\;ghosts}= 2\vec{n}^{2} +
\sum_{0<\vec{k}<\vec{n}} \frac{1}{4|\vec{n}|\vec{k}^{2}(\vec{n}
-\vec{k})^{2}}\nonumber \\
&&\times\left\{2|\vec{n}-\vec{k}|\vec{n}^{4}\vec{k}^{2}
-|\vec{n}-\vec{k}|\vec{n}^{2}\vec{k}^{4}
-|\vec{n}-\vec{k}|\vec{n}^{2}\vec{k}^{2}(\vec{n}\vec{k})
\right.\nonumber \\
&&+2|\vec{n}-\vec{k}|\vec{n}^{2}(\vec{n}\vec{k})^{2}
+2|\vec{n}-\vec{k}|\vec{k}^{4}(\vec{n}\vec{k})\nonumber \\
&&-3|\vec{n}-\vec{k}|\vec{k}^{2}(\vec{n}\vec{k})^{2}
-|\vec{n}-\vec{k}|(\vec{n}\vec{k})^{3}
+ 4\vec{n}^{4}|\vec{k}|(\vec{n}\vec{k})\nonumber \\
&&+\vec{n}^{2}|\vec{k}|^{5}-\vec{n}^{2}|\vec{k}|^{3}(\vec{n}\vec{k})
-6\vec{n}^{2}|\vec{k}|(\vec{n}\vec{k})^{2}\nonumber \\
&&\left.-2|\vec{k}|^{5}(\vec{n}\vec{k})
+5|\vec{k}|^{3}(\vec{n}\vec{k})^{2}
-|\vec{k}|(\vec{n}\vec{k})^{3}\right\}\nonumber \\
&&+\sum_{\vec{k}>0}\frac{1}{2|\vec{n}|\vec{k}^{2}(\vec{n}
+\vec{k})^{2}}\nonumber \\
&&\times\left\{2|\vec{n}+\vec{k}|\vec{n}^{4}\vec{k}^{2}
-|\vec{n}+\vec{k}|\vec{n}^{2}\vec{k}^{4}
+|\vec{n}+\vec{k}|\vec{n}^{2}\vec{k}^{2}(\vec{n}\vec{k})
\right.\nonumber \\
&&+2|\vec{n}+\vec{k}|\vec{n}^{2}(\vec{n}\vec{k})^{2}
-2|\vec{n}+\vec{k}|\vec{k}^{4}(\vec{n}\vec{k})\nonumber \\
&&-3|\vec{n}+\vec{k}|\vec{k}^{2}(\vec{n}\vec{k})^{2}
+|\vec{n}+\vec{k}|(\vec{n}\vec{k})^{3}
- 4\vec{n}^{4}|\vec{k}|(\vec{n}\vec{k})\nonumber \\
&&+\vec{n}^{2}|\vec{k}|^{5}+\vec{n}^{2}|\vec{k}|^{3}(\vec{n}\vec{k})
-6\vec{n}^{2}|\vec{k}|(\vec{n}\vec{k})^{2}\nonumber \\
&&\left.+2|\vec{k}|^{5}(\vec{n}\vec{k})
+5|\vec{k}|^{3}(\vec{n}\vec{k})^{2}
+|\vec{k}|(\vec{n}\vec{k})^{3}\right\}.
\label{central-ghosts2}
\end{eqnarray}.
 
Going in Eq. (\ref{central-ghosts2}) to the ``last-element limit''
we come to the following expression
\begin{equation}
[L(\vec{n}),L(-\vec{n})]_{c.e.\;ghosts} =
2n^{2} + \sum_{k=1}^{n-1} (2n-k)(n+k) =
\frac{13}{6}n^{3} - \frac{1}{6}n.
\label{central-ghosts3}
\end{equation}
It is easy to see that the result (\ref{central-ghosts3}) coincides
with the well-known expression for the Virasoro algebra.

Now, equating the coefficient at $n^{3}$ with an analogous coefficient
in Eq. (\ref{central-matter}) we come to the following critical
relation between the dimensionality of space and the matter content
of the theory:
\begin{equation}
d + d_{V} (N-1) + d_{F} 2^{[(N-1)/2 - 1]}
= 30 +\frac{5}{2} (N-2)(N+1).
\label{formula}
\end{equation}

Thus, we may claim that this short relation eventually
represents a necessary
condition for a nilpotency of the BRST charge (\ref{BRST}) in the
first quantum approximation. As is seen it imposes some correlations
between number of scalars $d$, vectors $d_{V}$, spinors $d_{F}$ and
dimensionality of the space.

\section{Conclusions}

In the Conclusions let us summarize briefly the key points of
suggested construction and make some comments about the results.

We have suggested a treatment of closed cosmological models by the
harmonic expansion of gravity and matter canonical variables and the
constraints. As basis harmonics we have chosen eigenfunctions of
Laplacian on the maximally symmetric manifold of the topology under
consideration. In doing so the structure constants of the involution
relations of constraints algebra could be expressed via Clebsch--
Gordan coefficients of the corresponding symmetry group.
This provides the general background for the multipole expansion of
constrained Hamiltonian dynamics of closed cosmological models,
and their subsequent quantization.
In this
paper, however, we restricted the further analysis to the case of $N$ -
dimensional generalization of the stationary closed Bianchi - $I$
cosmological model ($N$ - torus). Except the technical simplicity,
this choice of the particular model was motivated by the fact that
it has the stationary classical solutions.

Besides the multipole expansion, the construction implies splitting
of the constraint set into the area-preserving diffeomorphism
generators, forming a closed subalgebra, and the Virasoro-like
generators.

Performing the quantization we have constructed the BRST operator,
defining the Wick ordering for the gravity, matter and Virasoro-like
ghost excitations and the Weyl one for zero-modes and for the ghosts related
to area-preserving diffeomorphisms. Our further objective was an
examination of the nilpotency condition for the BRST charge. Actually,
in the nilpotency equation $\Omega^{2} = 0$,
we have managed to calculate the constant part of the first quantum
correction to the coefficient quadratic in ghosts .
In doing so we applied the expansion of
constraints and structure functions in powers of the dimensionless
parameter $\kappa = \frac{l_{P}^{(N-1)/2}}{V^{(N-1)/2N}}$.
Finally, we derived from the nilpotency condition the relation
between the dimensionality of the space $N$ and the spectrum of
matter fields (\ref{formula}). Although, the restrictions imposed
by Eq. (\ref{formula}) do not predict the unique spectrum and
spatial dimensionality, one can observe some curious
consequences, for example: stationary Bianchi-$I$ closed cosmology
is inconsistent without matter at any positive integer $N$.
Notice, also, that in
$N = 3$ case, the admissible matter spectra do not contain
sufficient number of degrees of freedom to be compatible with the
Standard Model or its generalizations. It is not too disappointing,
because the stationary closed Bianchi-$I$ model itself could hardly
be expected to describe the observable cosmology. On the other hand,
the very existence of such a relation allows to hope that
analogous conditions, being deduced for more realistic models (with
non-stationarity, other topology, massive particles and, perhaps,
with an extended symmetry), can display a better correspondence with
the realistic particle physics.

One more (a little bit speculative) way of fine-tuning the number of
matter degrees of freedom to the dimensionality of space could be found
in a generalized Kaluza-Klein ideology. In this way,
Eq. (\ref{formula}) can be thought of
as valid for the fundamental multidimensional
theory. That provides much more opportunities for
the matter spectrum in
low-dimensional effective theory by means of a proper compactification.

While an opportunity of interplay between matter content of
the Universe and its geometrical characteristics is
 of certain
interest, we would like to think that the designed scheme of Hamiltonian
treatment and quantization of closed cosmological models may open
some other
promising prospects for studying various aspects of quantum cosmology.

\section*{Acknowledgments}

We are thankful to Professor I. A. Batalin for useful discussions.
We are indebted to Professor S. Randjbar-Daemi for kind hospitality
at High Energy Section of ICTP, that allowed us to complete this work.
A. K. was partially supported by RFBR via grant No 96-02-16220 and
RFBR-INTAS via grant No 644. S. L. was partially supported by RFBR
via grant No 96-01-00482 and INTAS via grant No 93-2058.

\end{document}